\documentclass[12pt,preprint]{aastex}

\usepackage{epsfig, array, amsmath, delarray}
\shorttitle{Relativistic stellar aberration..}
\shortauthors{Turyshev}

\begin{document}

\title{Relativistic stellar aberration 
 for the \\ Space Interferometry Mission}

\author{Slava G. Turyshev\altaffilmark{1}}
\affil{Jet Propulsion Laboratory, California Institute
of  Technology, Pasadena, CA 91109}


\begin{abstract} 
This paper analyses the  
relativistic stellar aberration requirements for the Space Interferometry
Mission (SIM).  We address the issue of general relativistic
deflection of light by the massive self-gravitating bodies.  Specifically, we
present estimates for corresponding  deflection angles due to the  monopole
components of the gravitational fields of a large number of celestial
bodies in the solar system.  We study the possibility of deriving an
additional navigational constraints from the need to correct for the 
gravitational bending of light that is traversing the solar system. It turns out
that positions of the outer planets  presently may not have a sufficient
accuracy for the precision astrometry. However, SIM may significantly
improve those simply as a by-product of its astrometric program. We also
consider influence of the higher gravitational multipoles,  notably  the
quadrupole and the octupole ones, on the gravitational bending of
light.  Thus, one will have to model and  account for their influence
while observing the sources of interest in the close proximity of some 
of the outer planets, notably the Jupiter and the Saturn.  Results presented here
are different from the ones obtained elsewhere by the fact that we specifically
account for the differential nature of the future SIM astrometric campaign  (e.g.
observations will be made over the instrument's  field of regard with the size of
15$^\circ$). This, in particular, lets us to obtain a more realistic estimate 
for the accuracy of determination of the parameterized post-Newtonian (PPN)
parameter
$\gamma$.  Thus, based on a very conservative assumptions, we conclude
that  accuracy of $\sigma_\gamma
\sim 10^{-5}$ is achievable in the experiments conducted in the solar
gravity field. 
\end{abstract}    

\keywords{astrometry, solar system, relativity, SIM}
 
\section{Introduction}

The last quarter of a century have changed the  status of general relativity 
from a purely theoretical discipline to a practically important science. 
Present accuracy of astronomical observations requires relativistic  
description of light  propagation as well as the relativistically  correct 
treatment of the dynamics of the extended celestial bodies. As a result,  
some of the leading static-field post-Newtonian  perturbations in the 
dynamics of the  planets, the Moon and artificial satellites have been 
included in the equations of motion, and in time and position transformation. 
Due to enormous progress in the accuracy of astronomical observations we 
must now study the possibility of taking into account 
the much smaller relativistic effects caused by the post-post-Newtonian 
corrections to the solar gravitational field  as well as the post-Newtonian 
contributions from the lunar and  planetary gravity. It is also well 
understood that effects due to  non-stationary behavior of the solar 
system gravitational field as well as its  deviation from spherical 
symmetry should be also considered.  

Recent advances in the accuracy of astrometric observations  
have demonstrated  importance of taking into account   
the relativistic effects introduced by the solar system's  
gravitational environment. It is known that the reduction of the Hipparcos
data has necessitated the  inclusion of stellar aberration up to the terms
of the second order in $v/c$, and the general relativistic treatment of 
light bending due to the gravitational field of the Sun (see
discussion in \cite{Perryman92}) and Earth (please refer to
analysis in \cite{Gould93}).  The prospect of new high precision astrometric
measurements from space  with the  Space Interferometry Mission,  will require 
inclusion of relativistic effects at the $(v/c)^3$ level
as shown in \cite{aberr_memo}.    
At the level of accuracy expected from SIM, even
more subtle gravitational effects on astrometry from within the
solar system will  start to become apparent, such as the monopole and the
quadrupole components of  the gravitational fields of the  planets and
the gravito-magnetic effects caused by their  motions and rotations. 
Thus,  the identification of all possible sources  of  `astrometric'
noise that may contribute to the future  SIM astrometric campaign, is
well justified.

This work organized as follows: Section
\ref{sec:mon_defl} discusses the influence of the relativistic
deflection of light by the monopole components of the gravitational fields
of the solar system's bodies.  We present the model and our estimates for the
most important  effects  that will be influencing 
astrometric observations  of a few microarcsecond ($\mu$as) accuracy, that will
be made from within the solar system.
Section \ref{sec:three_env} will specifically address three the 
most intense gravitational  environments in the solar system,
namely the solar deflection of light and the gravitational defections
in the vicinities of the Jupiter and  the  Earth.
In Section \ref{sec:constr} we will discuss the constraints derived from the
monopole deflection of light on the navigation of the spacecraft and the
accuracy of the solar system ephemerides.
We study a possibility of  improvement in accuracy of
determination of  PPN  parameter $\gamma$ via astrometric tests of
general relativity  in the solar system. 
In Section \ref{sec:high_multipoles} we will discuss the effects of the
gravitational deflection of light by the higher gravitational multipoles (both
mass and current ones) of some of the bodies in the solar system.  
We will conclude the paper with the discussion of the results obtained
and  our recommendations for future studies.

\section{Gravity Contributions to the  Local Astrometric Environment}
\label{sec:mon_defl}

Prediction of the gravitational deflection of light was one of the
first successes of   general relativity.  Since the first confirmation by the
Eddington expedition in 1919,  the effect of gravitational  deflection 
has been studied quite extensively and currently analysis of almost every
precise  astronomical measurement must take this effect  into account 
(see \cite{Modest96}). According to general relativity, the  light rays  
propagating  near a gravitating body are achromatically scattered by the 
curvature of the  space-time generated by the body's gravity field.  The
whole  trajectory of the light ray is bent towards the body by an angle
depending on the strength of the body's gravity.   The solar gravity
field produces the largest effect on  the light  traversing the solar
system. 

\begin{figure}[ht]
\noindent   \hskip 50pt
 \begin{center}
\noindent   
\psfig{figure=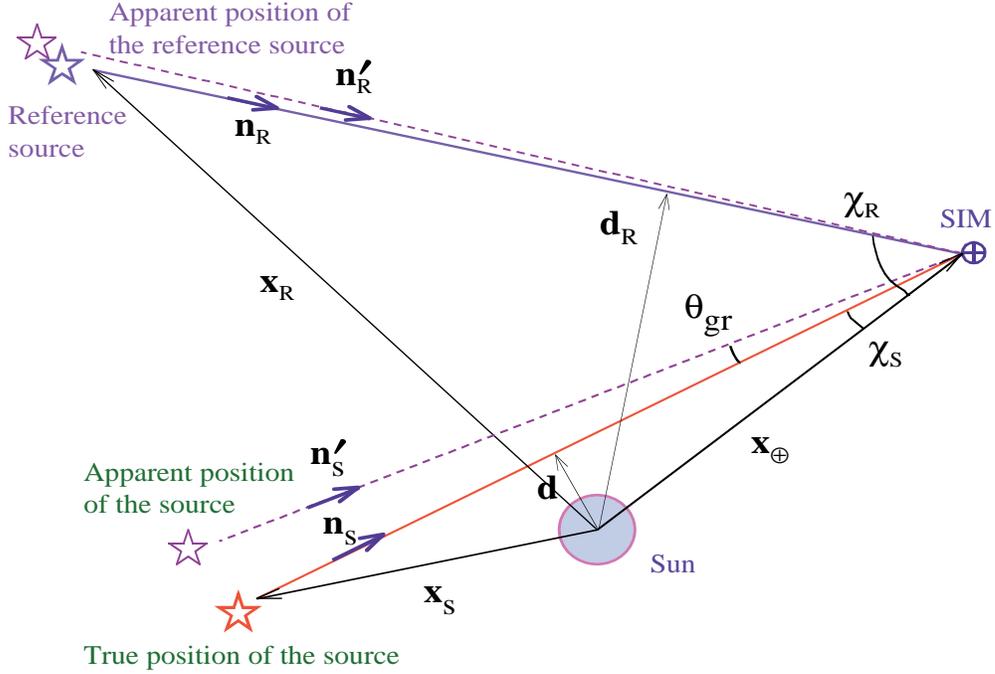,width=170mm,height=120mm} 
\vskip -45pt
    \caption{Geometry of gravitational deflection  of starlight by the 
Sun and the planets. 
      \label{fig:deflect}}
 \end{center}
\end{figure}

In the PPN formalism (please refer to \cite{Will93}), to first order
in the gravitational constant, $G$, the  solar deflection  angle 
$\theta^\odot_{\tt gr}$    depends  only on the solar  mass  
$M_\odot$ and the  impact parameter $d$  relative to the Sun:    
{}
\begin{equation}
\theta^\odot_{\tt gr}=\frac{1}{2}(\gamma+1){4GM_\odot\over c^2
d} ~\frac{1+\cos\chi}{2}.
\label{eq:deflec0}
\end{equation}
\noindent 
The star is assumed to be at a very large distance compared to the Sun,
and $\chi$ is the angular separation between the deflector and the
star. With the space observations  carried out by SIM, $\chi$ is not
necessarily a small angle.   The relevant geometry and notations are shown
in Fig. \ref{fig:deflect}. 


The absolute magnitude for the light deflection angle is maximal for the
rays grazing the solar photosphere, e.g. $\theta^\odot_{\tt
gr}=\frac{1}{2}(\gamma+1)\cdot1.751$ mas.  Most of the measurements  of
the gravitational deflection to date involved the solar gravity field,
planets in the solar system  or gravitational lenses. 
Thus, relativistic deflection  of light  has been observed,
with various degrees of precision,  on distance scales of
$10^9$  to $10^{21}$ m, and on mass scales from  $10^{-3}$  to $10^{13}$ 
solar masses, the upper ranges determined from the gravitational lensing of
quasars (\cite{Will93,Dar92,TreuhaftLowe91}).

The parameterized post-Newtonian (PPN) parameter $\gamma$ in the
expression (\ref{eq:deflec0}) represents the measure of the curvature of the
space-time  created by the unit rest mass (see \cite{Will93}). Note that general
relativity, when analyzed in  standard PPN  gauge, gives: $\gamma=1$. The
Brans-Dicke theory is  the most famous among the alternative theories of
gravity.  It contains, besides the metric tensor,  a scalar field $\phi$ 
and an arbitrary coupling constant $\omega$, related to this 
PPN parameter  as $\gamma= {1+\omega\over 2+\omega}$. 
The present limit  $|\gamma-1|\le3\times10^{-4}$, that was
recently obtained by  \cite{Eubanks97}, gives the constraint
$|\omega|>3300$.

In the Fig. \ref{fig:deflect} we emphasized the fact that the
difference of the apparent position of the source from it's true position 
depends on the impact parameter of the incoming light with respect to the
deflector.  For the astrometric accuracy of a few $\mu$as
and, in the case when the Sun is the deflector, positions of all observed
sources experiencing such an apparent displacement, except the ones
that are on  exactly opposite side from the instrument with respect
to the Sun,  e.g. $\chi=\pm\pi$. Indeed, the light rays coming from these
sources do not experience  gravitational deflection at all.
Thus, those observations may serve as an anchor to allow one to remove
the effects of the light bending from the high accuracy astrometric
catalogues. This is why, in order to correctly account for the effect of
gravitational deflection, it is important to process together the data
taken with the different separation angles from the deflector. SIM will
be observing the sky in a 15$^\circ$ patches of sky, as oppose to the
Very Long Baseline Interferometry (VLBI) that may simultaneously observe
sources with a much larger separations on the sky. To reflect this
difference, in our further estimations we will present results for the
two types of astrometric measurements, namely for the  absolute (single
ray deflection) and  differential  (two sources separated by  the
15$^\circ$ field of view) observations.

\subsection{Relativistic Deflection of Light by the Gravity Monopole}

In this Section we will address the question of how the
relativistic dynamics of  our  solar system will influence
the high-precision astrometric observations with SIM.
In particular, we will discuss the model, the parameterization of the 
quantities involved, as well as the physical meaning of the obtained
contributions. The main goal of this Section is to present a comparative analysis
of the  various  relativistic effects  whose presence must be  taken into
account in the modeling propagation of light through the solar system. In
particular we will concentrate on the effect of the relativistic deflection
of light traversing our solar system's internal gravitational environment.

\subsubsection{Modeling  the Astrometric Observations with an Interferometer}

The first step into a relativistic
modeling of the light path consists of determining the direction of the
incoming photon as measured by an observer located in the solar system as a
function of the barycentric coordinate position of the light source.  Apart
from second and third orders aberration the only other sizable effect is
linked to the bending of light rays in the gravitational field of solar
system bodies as shown by \cite{Turyshev98}. Effects of the gravitational
monopole  deflection of light are the largest among those in the solar system.

In order to properly describe this gravitational light-deflecting
phenomenon, one needs to define  the relativistic gravity geometric
contribution, $\ell_{\tt gr}$, to the  optical path difference  (OPD)
that is  measurable by  an interferometer in solar orbit. In a weak
gravity field approximation,  to first order in gravitational
constant $G$, an additional optical path difference introduced by the
gravitational bending (or, more specifically for the case of an
interferometer, the gravitational delay,
$\tau_{\tt gr}$,
see \cite{Sovers98}) of the electromagnetic signals, 
$\ell_{\tt gr}=c\tau_{\tt gr}$, takes the most simple and  
elegant form: 
{}
\begin{equation}
\ell_{\tt gr}= -(\gamma+1)\sum_B\frac{G}{c^2}\frac{M_B}{r_B}
\Big[{\vec{b}(\vec{s}+\vec{n}_B)\over 
1+(\vec{s}\cdot\vec{n}_B)}\Big], 
\label{eq:defl0}
\end{equation}

\noindent where $r_B$ is the distance from SIM to a
deflecting body $B$, $\vec{n}_B$  is the unit vector in
this direction.   Also $~\vec{b}=b\vec{n}$ ~and  $\vec{s}$ are  
the vector of interferometer's baseline   and  the  unit vector of the 
unperturbed direction to the source at  infinity correspondingly. 
Note that Eq.(\ref{eq:defl0}) is written in the approximation neglecting
the terms of the order $\sim M_Bb^2/r^2_B$, which could be reinstated, 
if needed.
 
\begin{figure}[ht]
 \begin{center}\vskip -15pt   
\noindent   \hskip -10pt   
\psfig{figure=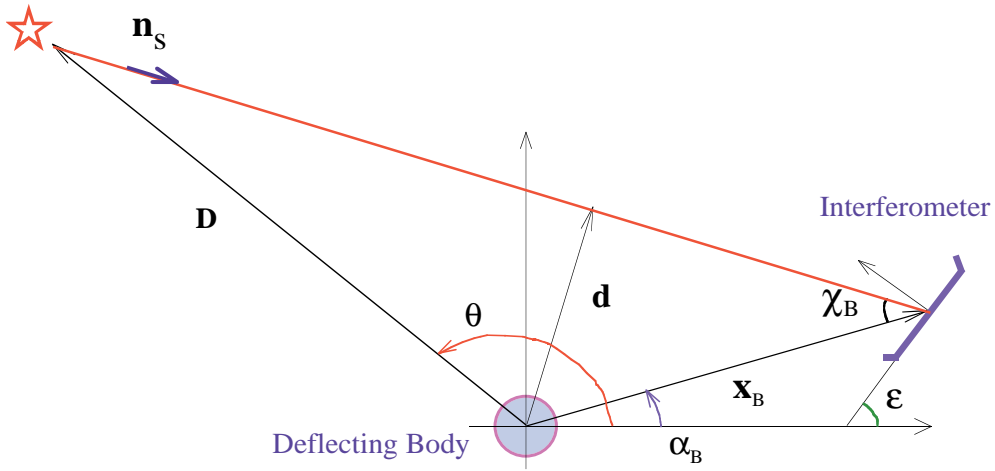,width=140mm,height=95mm} 
\vskip -20pt
    \caption{Parameterization and notations for the 
gravitational deflection of light. 
      \label{fig:defl_param}}
 \end{center}
\end{figure}

For the purposes of this study it is sufficient to confine our analysis 
to a planar motion and parameterize  the quantities involved  as
follows: 
\begin{equation}
\vec{b}= b \,(\cos\epsilon,\, \sin\epsilon), ~~~~~
\vec{r}_B= r_B\,(\cos\alpha_B, \,\sin\alpha_B), ~~~~~
\vec{s}= (\cos\theta,\, \sin\theta),
\end{equation}

\noindent where $\epsilon$ is the angle of the baseline's
orientation with respect to the instantaneous body-centric coordinate
frame,  $\alpha_B$ is the right assention angle of the interferometer
as seen from the this frame and $\theta$ is the direction to the
observed source correspondingly. The geometry of the problem and the
discussed notations  are presented  in the Figure \ref{fig:defl_param}.  
It is convenient to  express the gravitational contribution to the
total optical path difference Eq.(\ref{eq:defl0}) in terms of the deflector and
the source separation angle $\chi_B$   as observed by the interferometer.   
The necessary relation that expresses the source's position angle
$\theta$ via the  separation angle $\chi_B$ may be given as:
\begin{equation}
\theta=\pi+\alpha_B-\chi_B-\arcsin\Big[\,\frac{r_B}{D}\,
\sin\chi_B\,\Big].
\label{eq:theta_chi}
\end{equation}
\noindent 
As a result, we can now rewrite the gravitational deflection's
contribution to the total OPD, Eq.(\ref{eq:defl0}),  in the following
form: {}
\begin{equation}
\ell_{\tt gr}= -(\gamma+1)\sum_B\frac{G}{c^2}\frac{M_Bb}{r_B}
\Big[\cos(\epsilon-\alpha_B)+\sin(\epsilon-\alpha_B)
\frac{1+\cos\chi^*_B}{\sin\chi^*_B}\Big],
\label{eq:defl_chi}
\end{equation}
\noindent where $\chi^*_B= \chi_B+\arcsin\big[\,\frac{r_B}{D}\,
\sin\chi_B\,\big]$.
We can further assume that the source is located at a very
large distance, $D$, compare to the distance between the interferometer
and the deflector, so that the following inequality holds 
$r_B\ll D$ for every body in the solar system. This allows  us to 
neglect the presence of the last term in the equation 
Eq.(\ref{eq:theta_chi}) for the estimation purposes only.  
Also a complete analysis
of  phenomenon of the gravitational deflection of light will have to account for
the time dependency in all the quantities involved. Thus, one will have to use
the knowledge of the position of the spacecraft in the solar system's
barycentric reference frame, the instrument's orientation in the proper
coordinate frame, the time that was spent in a particular orientation, the
history of all the maneuvers and re-pointings of the instrument, etc.  These
issues are closely related to the principles of the operational mode of the
instrument that is currently still being developed.  

\subsubsection{Modeling for the Absolute Astrometric Observations}

Equation Eq.(\ref{eq:defl_chi}) represents the fact that the gravitational
field is affecting the propagation of the  electromagnetic signals in a two ways,
namely by delaying them and by deflecting the light's trajectory from the
rectilinear one.    Thus, the first term in the square brackets on the right
hand-side of this equation  is the term that describes
the gravitational delay of the infallen electromagnetic signal.
This term is independent on the source's position on the sky and depends only on 
the orientation of the baseline vector and the direction to the deflector. More
precisely, it depends on the gravity generated by the body at the
interferometer's location and the  projected  baseline vector onto the
direction to the source 
$(\vec{b}\cdot\vec{n}_B)=b\,\cos(\epsilon-\alpha_B)$. For the purposes of this
study it is sufficient to discuss  only the magnitude of this effect in
terms of its contribution to the  astrometric measurement:  
{}
\begin{equation}
\theta^{\tt delay}_{\tt gr}=\frac{\ell^{\tt delay}_{\tt gr}}{b}=
-(\gamma+1)\sum_B\frac{G}{c^2}\frac{M_B}{r_B}.
\label{eq:delay_grav}
\end{equation}
\noindent

The second  term in the equation Eq.(\ref{eq:defl_chi}) is responsible for the
relativistic deflection of light and will be the main topic of our
further discussion.    In our future analysis  we
will be interested only in   magnitudes of the angles of relativistic deflection
of light, so it is convenient to choose [only for the estimation purposes!] such
an orientation of the baseline vector (e.g. angle
$\epsilon$)  and the vector of mutual orientation of the instrument and the 
deflecting body (e.g. angle $\alpha_B$)  that maximizes the effect of
the gravitational deflection of light. By choosing the orientation angles as 
$\epsilon-\alpha_B=\frac{\pi}{2}$, we can are neglect it's presence.  This
allows one to concentrate only on the phenomenon of the gravitational deflection
and to re-write  the  contribution of this effect to  the total OPD,
Eq.(\ref{eq:defl_chi}),   as $\ell_{\tt gr}=-\sum_B\ell_{\tt gr}^B,$ 
with   individual contributions $\ell_{\tt gr}^B$ given  by
\begin{equation}
\ell^B_{\tt gr}=
(\gamma+1)~\frac{G}{c^2}\frac{M_B b}{r_B}~
\frac{1+\cos\chi_B}{\sin\chi_B}.
\label{eq:b_opd}
\end{equation}

\noindent It is also convenient express this
additional OPD in terms of the corresponding deflection  angles
$\theta^B_{\tt gr}$, which simply have the form:
{} 
\begin{equation}
\theta^B_{\tt gr}=\frac{\ell^B_{\tt gr}}{b}=
(\gamma+1)~\frac{G}{c^2}\frac{M_B}{r_B}~\frac{1+\cos\chi_B}{\sin\chi_B}.
\label{eq:b_defl}
\end{equation} 
\noindent Note that $~r_B\sin\chi_B=d_B$ is the impact
parameter of the incident light ray with respect to a particular deflector
as seen by the interferometer. By substituting this result into the 
formula (\ref{eq:b_opd}) one obtains expression similar to that
given by Eq.(\ref{eq:deflec0}). 

The obtained expressions Eqs.(\ref{eq:b_opd})-(\ref{eq:b_defl}) are most
appropriate to estimate the magnitude of the gravitational bending effects
introduced into  absolute astrometric measurements. They are useful
in understanding the ``asymptotic value'' of the effect for a large number
of observations, $N\gg1$. However, one needs an additional set of 
equations suitable to describe the accuracy of  measurements
during   differential astrometry studies with SIM. 

\subsubsection{Differential Astrometric Measurements}

The necessary  expression for the differential OPD may be  simply
obtained by subtracting OPDs for the different sources one from one
another. This procedure resulted in  the following expression:   {} 
\begin{equation}
\delta\ell^B_{\tt gr} = \ell^B_{1\tt gr}-\ell^B_{2\tt gr}=
-(\gamma+1) \sum_B\frac{G}{c^2}\frac{M_B}{r_B}
\Big[\frac{\vec{b}(\vec{s}_1+\vec{n}_B)}
{1+(\vec{s}_1\vec{n}_B)}-
\frac{\vec{b}(\vec{s}_2+\vec{n}_B)}
{1+(\vec{s}_2\vec{n}_B)}\Big],
\label{eq:delay}
\end{equation}
 
\noindent where ${\vec s}_1$ and $\vec{s}_2~$ are the  barycentric
positions of the primary and the secondary objects. By using   
parameterization for the  quantities involved similar to that  above,
this expression may be presented in terms of the  deflector-source
separation angles,
$\chi_{1B}, ~\chi_{2B}$,  as follows:
\begin{equation}
\delta\ell_{\tt gr}= -(\gamma+1)\sum_B\frac{G}{c^2}\frac{M_Bb}{r_B}
\sin(\epsilon-\alpha_B)\,
 \,\frac{\sin\frac{1}{2}(\chi_{B2}-\chi_{B1})}
{\sin\frac{1}{2}\chi_{B1}\sin\frac{1}{2}\chi_{B2}}. 
\label{eq:defl_chi_diff}
\end{equation}

\noindent The purpose of this was was only to estimate the influence
of the solar system's gravity field on the propagation of light.
We will  concentrate  on obtaining the magnitudes
of the deflection angles only and will not try to reconstruct the 
complicated functional dependency of the effect on the number of mutual
orientation angles.  This allows us use the expression
Eq.(\ref{eq:defl_chi_diff}) with such an   orientation between the
baseline vector, $\epsilon$,  and deflector$_-$instrument  angle,
$\alpha_B$, that maximizes  contribution of each individual deflector
for a particular orbital position of the spacecraft.
As a result, we may well require that 
$\epsilon-\alpha_B=\frac{\pi}{2}$ and expression
Eq.(\ref{eq:defl_chi_diff}) may be re-written as 
$\delta\ell_{\tt gr}=-\sum_B \,\delta\ell_{\tt
gr}^B$, with the individual contributions $\delta\ell_{\tt gr}^B$  
having the form
\begin{equation}
\delta\ell^B_{\tt gr}=
(\gamma+1)~\frac{G}{c^2}\frac{M_B b}{r_B}~
\frac{\sin\frac{1}{2}(\chi_{2B}-\chi_{1B})}
{\sin\frac{1}{2}\chi_{1B}\cdot\sin\frac{1}{2}\chi_{2B}}.
\label{eq:diff_opd}
\end{equation}

Finally, it is convenient to  express this
result for $\delta\ell^B_{\tt gr}$ in terms of the corresponding
deflection  angle $\delta\theta_{\tt gr}$. Similarly to the expression 
Eq.(\ref{eq:b_defl}), one obtains:  
{}
\begin{equation}
\delta\theta^B_{\tt gr}=\frac{\delta\ell^B_{\tt gr}}{b}=
(\gamma+1)~\frac{G}{c^2}\frac{M_B}{r_B}~
\frac{\sin\frac{1}{2}(\chi_{2B}-\chi_{1B})}
{\sin\frac{1}{2}\chi_{1B}\cdot\sin\frac{1}{2}\chi_{2B}}. 
\label{eq:diff_deflec}
\end{equation}

\subsubsection{Deflection of Grazing Rays by the Bodies of the Solar
System}
In this section we will obtain the estimates for the effects that characterize
the intensity of the gravitational environment in the solar system.
The most natural and convenient way to do that is to discuss
the magnitudes of the angles of the gravitational deflection of light rays that
grazing the surfaces of the celestial bodies.

                              
\begin{table} 
\begin{center}
\caption{Relativistic monopole deflection of grazing 
(e.g. $\chi_{1B}={\cal R}_B$) light rays by the bodies of the solar system at the
SIM's location that is assumed to be placed in the solar Earth-trailing  orbit.
For the differential observations the two stars are assumed to be separated by
the size of the instrument's field of regard. For the grazing rays,   position
of the primary star is assumed to be on the limb of the deflector. Moreover,
results are given for the smallest distances from SIM to the bodies (e.g.when the
gravitational deflection effect is largest). For the Earth-Moon system we took
the SIM's position at the end of the first half of the first year mission at
the distance of 0.05 AU from the Earth.  Presented in the right column of this
Table are the magnitudes of the body's individual contributions to
the  gravitational  delay of  light  at the SIM's location  (note
that it is unobservable in the case of differential astrometry with SIM).} 
\label{tab:mon}
\vskip 10pt 
\begin{tabular}{|r|r|r|r|r||c|} \hline

Solar  & Angular size   & 
\multicolumn{3}{c||}{Deflection of grazing rays} & Delay
\\[2pt]\cline{3-5}       
system's & from SIM,  &  absolute  &
diff. $[15^\circ]$ & diff. $[1^\circ]$ & $\theta^{\tt delay}_{{\tt gr}B}$,\\[2pt]
object & ${\cal R}_B$, arcsec &  $\theta^B_{\tt gr}, \mu$as  &
$\delta\theta^B_{\tt gr},~\mu$as & 
$\delta\theta^B_{\tt gr}, ~\mu$as & $\mu$as\\[2pt]\hline\hline

Sun      & 0$^\circ$.26656 & 1$''$.75064   
         & 1$''$.72025 & 1$''$.38221 & 4072.29\\

Sun at 45$^\circ$& 45$^\circ$ & 9,831.39  
         & 2,777.97 &  237.66 & -same-\\

Moon     & 47.92690 & 25.91 &  25.87 & 25.56 & 0.003\\

Mercury  & 5.48682  & 82.93  & 82.92  & 82.81 & 0.001\\

Venus    & 30.15040 & 492.97  & 492.69 &  488.88 & 0.036\\

Earth    & 175.88401 & 573.75  & 571.90 & 547.03 & 0.245\\
 
Mars     & 8.93571 & 115.85 & 115.83 & 115.57 & 0.003\\

Jupiter  & 23.23850 & 16,419.61   & 16,412.60 & 16,314.30 & 0.925\\

Jupiter at 30$''$ & 30.0 & 12,719.12   & 12,712.03 & 12,614.21 & -same-\\

Saturn   & 9.64159 & 5,805.31 & 5,804.27  & 5,789.79 & 0.126\\

Uranus   & 1.86211 & 2,171.38  & 2,171.30 & 2,170.26 & 0.010\\

Neptune  & 1.18527 & 2,500.35  & 2,500.29  & 2,499.52 & 0.007\\
 
Pluto    & 0.11478 & 2.82  & 2.82 & 2.82 & 0.00\\\hline

\end{tabular} 
\end{center} 
\end{table}

 In the two previous sections we have obtained expressions suitable to
describe  effects of the gravitational bending of light on both
absolute and differential astrometric observations. Now we have all
that is necessary  to estimate the influence of the solar
system's gravity field on the future high-accuracy astrometric
observations.  The corresponding post-Newtonian effects for grazing 
rays, deflected by the  solar system's bodies, are given  in the Table
\ref{tab:mon}.  To obtain these estimates we  used  the
physical constants and the solar system's parameters  
that are given in the Tables \ref{tab:solar_system_param} and
\ref{tab:constants}. The results presented in the terms of the following 
quantities: 
\begin{itemize}
\item[i).] for the absolute astrometry   we  present the results 
in terms of the absolute measurements  
$\ell^B_{\tt gr}, \theta^B_{\tt gr}$ defined by  
Eq.(\ref{eq:b_opd}) and Eq.(\ref{eq:b_defl}); 
\item[ii).] to describe the
differential observations we  use the relations Eq.(\ref{eq:diff_opd})
and Eq.(\ref{eq:diff_deflec}) and  
express  those in terms  of the differential astrometric observables,
namely   $~\delta\ell^B_{\tt gr}, ~\delta\theta^B_{\tt gr}$. 
\end{itemize}
Note, that the angular separation of the
secondary star will always be taken larger than that for the primary. It
is convenient to study the case of the most distant available
separations of the sources. In the case of SIM, this is the  size
of the  field of regard (FoR). Thus for the wide-angle astrometry the
size of  FoR will be $15^\circ\equiv\frac{\pi}{12}$, thus
$\chi_{2B}=\chi_{1B}+\frac{\pi}{12}$. For the
narrow-angle observations this size is 
FoR $=1^\circ\equiv\frac{\pi}{180}$, thus for this type of  astrometric
observations we will use
$\chi_{2B}=\chi_{1B}+\frac{\pi}{180}$.  Additionally, the baseline
length will be assumed $b=10$ m. 

In the Table \ref{tab:mon} we also presented the magnitudes of 
the individual solar and   planetary contributions   to the total
gravitational  delay of  light  traversing the solar system at the SIM's
location, $\theta^{\tt delay}_{{\tt gr}B}$. This contribution is given by  
Eq.(\ref{eq:delay_grav}) and  it affects only the absolute astrometric
measurements. Thus, one may see that it is important to account for this effect
only in case of gravity contributions of the Sun and the Jupiter only.

\subsection{Critical Impact Parameter for High Accuracy Astrometry}
 
The estimates, presented in the Table \ref{tab:mon} have
demonstrated that it is very important to correctly model and account
for gravitational influence of the bodies of the
solar system. Depending on the impact parameter $d_B$ (or planet-source separation
angle, $\chi_B$), one will have  to account for the   post-Newtonian
deflection of  light  by   a  particular planet. 
Most important is that one will have to permanently monitor the
presence of some of the bodies of the solar system during all
astrometric observations, independently on the position of the
spacecraft in it's solar orbit and the observing direction. The 
bodies that introduce the biggest astrometric inhomogeneity
are the Sun, the Jupiter and the Earth (especially at the beginning of
the mission, when the spacecraft is in the Earth' immediate proximity).

Let us introduce a measure of such a gravitational inhomogeneity
due to a particular body in the solar system.
To do this,  suppose that future astrometric experiments with 
SIM  will be capable to measure astrometric parameters   
with  accuracy of $\Delta \theta_0=\Delta k
~\mu$as, where $\Delta k$ is some number characterizing the accuracy of
the instrument [e.g. for a single measurement accuracy $\Delta k =8$ 
and for the mission accuracy $\Delta k=4$]. 
Then, there will be a critical distance from the body,   beginning from
which, it is important to account for the presence of the body's gravity
in the vicinity of the observed part of the sky. Let's call this
distance |  critical impact parameter,
$d^B_{\tt crit}$, the closest distance between the body and the light
ray that is gravitationally deflected to the  angle 
\begin{equation}
\theta^{\tt c}_{\tt gr}(d^B_{\tt crit}) = \Delta\theta_0 = \Delta k
~\mu{\sf as}.
\label{eq:crit_condition}  
\end{equation}
\noindent
The necessary expression for $d^B_c$ may be obtained with the help of
Eq.(\ref{eq:b_defl}) as follows:    
\begin{equation}
d^B_{\tt crit} = \pm\,\frac{4\,\mu_B}{\Delta\theta_0}  
 \Big[\,1+\Big(\frac{2\mu_B}{\Delta\theta_0\,r_B}\,\Big)^2
\Big]^{-1},
\label{eq:crit_dist}
\end{equation}
\noindent where $\mu_B=c^{-2}GM_B$ is the usual notation for the
gravitational ($\frac{1}{2}$Schwarzshild) radius of the body.
The choice of the sign depends on the relation between the terms, thus
if  $4\mu_B/r_B >\Delta\theta_0$, then the negative sign
should be chosen. The negative sign reflects the fact that the impact
parameter becomes critical [e.g.  satisfies the equation
Eq.(\ref{eq:crit_condition} )] for the  sources that
have the deflector-source separation angle on the sky
$|\chi_B|$ more than $90^\circ$.  This is definitely true for the
case of accounting for gravitational influence of the two solar system
bodies, namely the Sun and the Jupiter. For the other bodies of the
solar system the ratio holds as ~
$4\mu_B/r_B \ll \Delta\theta_0=$ few $\mu$as, thus
significantly simplifying the analytical expression
Eq.(\ref{eq:crit_dist}).

The formula for the critical distance, Eq.(\ref{eq:crit_dist}), may be
given in a slightly different form, representing  the critical angles, 
$\alpha_c^B$, that correspond to this critical distance from the body:
\begin{equation}
\alpha_c^B=\arcsin\big(\,\frac{d^B_{\tt crit}}{r_B}\,\big) =
\arcsin\Big[\,\frac{4\,\mu_B}{\Delta\theta_0\,r_B}  
 \big[\,1+\Big(\frac{2\mu_B}{\Delta\theta_0\,r_B}\,\Big)^2
\,\big]^{-1}\Big].
\label{eq:crit_angle}
\end{equation}
\noindent Different forms of the critical impact parameters
$d^B_{\tt crit}$ for $\Delta k=1$ are given in the 
Table \ref{tab:crit_dist}. With the help of  
Eq.(\ref{eq:crit_dist}), the results given in
this table are easily scaled for any astrometric accuracy $\Delta k$.


\begin{table}
\begin{center}
\caption{Relativistic monopole deflection of light: the angles
and the critical distances for $\Delta\theta_0=1~\mu$as astrometric
accuracy. Negative critical distance for the Sun represents the fact
that the Sun-source critical separation angle, $\alpha_c^B$, is larger
than 90$^\circ$. To visualize the solar gravitational  deflection
power  note that a light ray coming perpendicular to the ecliptic plane
at the distance  of 
$d_{1\,\mu{\sf as}}=$ 4072.3 AU from the Sun will be still deflected
by the solar gravity to  $1~\mu{\sf as}$! The critical distances 
for the Earth are given for two distances, namely for 0.05 AU 
(27$^\circ$.49) and 0.01 AU (78$^\circ$.54).
\label{tab:crit_dist}}
\vskip 10pt 
\begin{tabular}{|r|c|r|c|c|} \hline

Object & $\theta^B_{\tt gr}, ~\mu$as &
\multicolumn{3}{c|}{Critical distances 
for  accuracy of ~$1~\mu$as}
\\[2pt]\cline{3-5}   
  &   & $d^B_{\tt crit}$, ~cm & $d^B_{\tt crit}$, ~deg & 
$d^B_{\tt crit}$, ~${\cal R}_B$\\\hline\hline

Sun     & 1$''$.75064    &   
          $-7.347 \times 10^9$ & $\pi+ 101''.3$& 
          $0.11 \cdot {\cal R}_\odot$\\

Moon    &  25.91  & $4.501 \times 10^9$ & 
           $0^\circ.34 - 1^\circ.72$ & $25.9 \cdot {\cal R}_m$\\

Mercury & 82.93  & $2.023 \times 10^{10}$ & 
          $0^\circ.06 - 0^\circ.13$ &
          $82.9\cdot {\cal R}_{Me}$\\

Venus   & 492.97  & $2.982 \times 10^{11}$ &
          $0^\circ.66 - 4^\circ.13$  & $492.9\cdot {\cal R}_V$\\

Earth   & 573.75 & $3.453 \times 10^{11}$ & 
          27$^\circ.49-78^\circ$.54 &  
          $ 541.4\cdot {\cal R}_\oplus$\\
 
Mars    & 115.85 & $3.931\times 10^{10}$ &
           0$^\circ.06-0^\circ$.29 & $115.9\cdot {\cal R}_{Ms}$\\

Jupiter & 16,419.61 & $6.270\times 10^{13}$ & 
          $64^\circ.06-88^\circ.51$ & 
          $8,849 \cdot {\cal R}_J$\\
Saturn  & 5,805.31  & $3.420 \times 10^{13}$ & 
          $12^\circ.56-15^\circ.45$ & 
          $5,700 \cdot {\cal R}_S$ \\

Uranus  & 2,171.38 & $5.319\times 10^{12}$ & $1^\circ.01-1^\circ.12$ & 
          $2,171\cdot {\cal R}_U$ \\

Neptune  & 2,500.35 & $6.276 \times 10^{12}$ & 
                          $0^\circ.77-0^\circ.82$  & 
           $2,500\cdot {\cal R}_N$ \\
 
Pluto    & 2.82  & $9.025\times 10^8$ & $0''.31-0''.32 $  
         &  $2.8 \cdot {\cal R}_P$\\\hline

\end{tabular} 
\end{center} 
\end{table}

\begin{table}
\begin{center}
\caption{Relativistic deflection of  light by some
 planetary satellites.}\label{tab8} 
\vskip 10pt 
\begin{tabular}{|r|c|c|c|r|r|r|} \hline

Object &  Mass,   & Radius,   & Angular size,  &
Grazing & 
\multicolumn{2}{c|}{1 $\mu$as critical radius}\\\cline{6-7}
&    $10^{25} $~g &  ${\cal R}_B$, ~km &  ${\cal R}_B$, arcsec &
$\theta^B_{\tt gr}, ~\mu$as &  $d_{\tt crit}, $~km &  $d_{\tt crit},$ 
${\cal R}_{planet}$\\\hline\hline

Io    & 7.23 & 1,738 & 0.570056 & 25.48 & 44,291
            &  $0.63\cdot{\cal R}_J$ \\  
Europa  & 4.7 & 1,620 & 0.531353  & 17.77
        & 28,793 & $0.41\cdot{\cal R}_J$ \\
Ganymede  & 15.5 & 1,415 & 0.464114 & 67.11 & 94,954
          & $1.34\cdot{\cal R}_J$ \\  
Callisto  & 9.66 & 2,450 & 0.803589& 24.15 
          &59,178& $0.84\cdot{\cal R}_J$ \\  
Rhea  & 0.227 & 675 & 0.108468 & 2.06 & 1,391  
      & $0.02\cdot{\cal R}_S$ \\  
Titan  & 14.1 & 2,475 & 0.397715 & 34.90  & 86,378 
       &$1.44\cdot{\cal R}_S$ \\  
Triton  & 13 & 1,750 & 0.082638 & 45.51 & 79,639
        &$3.17\cdot{\cal R}_N$ \\ \hline

\end{tabular} 
\end{center} 
\end{table}

\subsection{Deflection of Light by Planetary Satellites}

One may expect that the planetary satellites will affect the astrometric
studies a light ray would pass in their vicinities.  Just for
completeness of our study we would like to present the estimates for the
gravitational deflection of light by the  planetary satellites and the
small bodies in the solar system. The corresponding 
estimates for deflection angles,
$\theta^B_{\tt gr}$, and   critical distances, $d_{\tt crit}$ are
presented  in the Table \ref{tab8}. Due, to the fact that the angular
sizes for those bodies are much less than the smallest field of regard
of the SIM instrument (e.g. FoR=1$^\circ$), the results for the
differential observations will be effectively insensitive to the size
of the the two available FoRs. The obtained results  demonstrate the fact that 
observations of of these objects with that size of FoR will evidently
have the effect from the relativistic bending of light. Thus, we have
presented there only the angle for the absolute gravitational deflection
in terms of quantities $\theta^B_{\tt gr}$.  As a result, the
major satellites of Jupiter, Saturn and Neptune should also be included
in the model   if the light ray  passes close to these bodies.

\subsection{Gravitational Influence of Small Bodies}

Additionally,  for   $\Delta k ~\mu{\rm as}$ astrometric accuracy,  one
needs to  account for the post-Newtonian deflection of  light due to  
rather a large number of small bodies in the solar system having a mean
radius  {}
\begin{equation}
 {\cal R}_B \ge  624 ~
\sqrt{\Delta k\over \rho_B}\hskip 8pt ~{\sf km}. 
\label{(2.11)}
\end{equation}

The deflection angle for the largest asteroids Ceres, Pallas and Vesta 
for $\Delta k=1$ are given in the Table \ref{tab7}.
The quoted properties of the asteroids were taken from  
\cite{StandishHellings89}.
Positions of these asteroids are  known and they  are incorporated 
in the JPL ephemerides. However, due to the fact that the other small
bodies (e.g. asteroids, Kuiper belt objects, etc.)
may produce a stochastic noise in the future
astrometric observations with SIM, so they should also be properly
modeled. 
\begin{table}[h]
\begin{center}
\caption{Relativistic deflection of light by the asteroids. 
\label{tab7}} 
\vskip 10pt
\begin{tabular}{|r|c|c|c|} \hline

Object   & $\rho_B,$ ~g/cm$^3$
& Radius,  km & $\theta^B_{\tt gr}, ~\mu$as \\\hline\hline

Ceres       & 2.3& 470 &1.3 \\ 
Pallas      & 3.4 & 269 & 0.6 \\ 
Vesta       & 3.6 & 263 & 0.6 \\ 
Class S     & 2.1 $\pm$ 0.2 & TBD & $\le 0.3$\\ 
Class C     & 1.7 $\pm$ 0.5 & TBD & $\le 0.3$\\\hline 

\end{tabular} 
\end{center} 
\end{table}

\section{Most Gravitationally Intense Astrometric Regions for SIM}
\label{sec:three_env}

The properties of the solar system's gravity field presented in the
Tables \ref{tab:mon} and \ref{tab:crit_dist} suggesting that the most
intense gravitational environments   in the solar system are those
offered by the Sun and two planets, namely the Earth and the
Jupiter.  In this Section we will analyze these
regions in  a more details. 

\subsection{Gravitational Deflection of Light by the Sun} 

From the  expressions Eq.(\ref{eq:b_defl}) and 
Eq.(\ref{eq:diff_deflec}) we obtain the relations for relativistic
deflection of light by the solar gravitational monopole. The expression
for the   absolute astrometry takes the form:   
{} 
\begin{equation}
\theta^\odot_{\tt gr}= (\gamma+1)~\frac{G}{c^2}
\frac{M_\odot}{r_\odot}~\frac{1+\cos\chi_{1\odot}}{\sin\chi_{1\odot}}=
4.072\cdot\frac{1+\cos\chi_{1\odot}}{\sin\chi_{1\odot}} ~~~~{\sf mas},
\label{eq:sun_defl}
\end{equation}
\noindent where $\chi_{1\odot}$ is the Sun-source
separation angle,  $r_\odot=1$ AU, and $\gamma=1$.
Similarly, for differential  astrometric observations  one obtains:
\begin{equation}
\delta\theta^\odot_{\tt gr}= (\gamma+1)\frac{G}{c^2} ~
\frac{M_\odot}{r_\odot}\frac{\sin\frac{1}{2}(\chi_{2\odot}-\chi_{1\odot})}
{\sin\frac{1}{2}\chi_{1\odot}\cdot\sin\frac{1}{2}\chi_{2\odot}} 
=4.072\cdot\frac{\sin\frac{1}{2}(\chi_{2\odot}-\chi_{1\odot})}
{\sin\frac{1}{2}\chi_{1\odot}\cdot\sin\frac{1}{2}\chi_{2\odot}}~~~~
{\sf mas}, 
\label{eq:sun_def_defl}
\end{equation}
\noindent with $\chi_{1\odot}, \chi_{2\odot}$ being the Sun-source
separation angles for the primary and the secondary stars
correspondingly. Remember that we use the two stars separated by 
the SIM's field of regard, namely 
$\chi_{2\odot}=\chi_{1\odot}+\frac{\pi}{12}$.  The solar angular dimensions
from the Earth' orbit are  calculated to be  
${\cal R}_\odot=0^\circ.26656$. 
This angle corresponds to a deflection of light to $1.75065 $~arcsec 
on the limb of the Sun.  Results for the most interesting  range of
$\chi_{1\odot}$   are given in the Table
\ref{tab:sun}.

\begin{table}
\begin{center}
\caption{Magnitudes of the gravitational deflection angle vs. the
Sun-source separation angle $\chi_{1\odot}$.}
\label{tab:sun}
\end{center}
\vskip -10pt
\begin{tabular}{|r|c|c|c|c|c|c|c|}
\hline

\tt Solar    & \multicolumn{7}{c|}{small $\chi_{1\odot},$ ~deg}\\[1pt]
\cline{2-8}
\tt deflection  &$0^\circ.27$ &$0^\circ.5$&$1^\circ$&$2^\circ$&
$5^\circ$&$10^\circ$&$15^\circ$\\\hline  
$~\theta^\odot_{\tt gr}$, ~mas &
1,728
& 
933.295 & 
466.639 &
233.302 & 
93.271 & 
46.547 & 
30.932 \\[3pt] \hline
$\delta\theta^\odot_{\tt gr}~[15^\circ]$, ~mas &
1,698 &
903.372& 
437.663 &
206.053 &
70.176 &
28.178 & 
15.734 \\[3pt] \hline
$\delta\theta^\odot_{\tt gr}~[1^\circ]$, ~mas &
1,361 &
622.212& 
233.337 &
77.787 &
15.567 &
4.254 & 
1.956 \\[3pt] \hline
\end{tabular} 
\vskip 1pt
\begin{tabular}{|r|c|c|c|c|c|c|c|c|}
\hline
\tt Solar    & \multicolumn{8}{c|}{large $\chi_{1\odot},$ ~deg }  
\\[1pt]\cline{2-9} 
\tt deflection & $20^\circ$ & $40^\circ$ & $45^\circ$ & $50^\circ$
& $60^\circ$ & $70^\circ$ & $80^\circ$ & $90^\circ$\\\hline 
$~\theta^\odot_{\tt gr}$, ~mas &
23.095&
11.189 & 
9.832 &
8.733 &
7.053 &
5.816 &
4.853 &
4.072 \\[3pt] \hline
$\delta\theta^\odot_{\tt gr}~[15^\circ]$, ~mas &
10.180&
3.366 & 
2.778 &
2.341 &
1.746 &
1.372 &
1.122 &
0.948 \\[3pt] \hline
$\delta\theta^\odot_{\tt gr}~[1^\circ]$, ~mas &
1.123 &
0.297 & 
0.238 &
0.195 &
0.140 &
0.107 &
0.085 &
0.071 \\[3pt] \hline
\end{tabular} 
\end{table}
\begin{figure}
\noindent
\vskip -50pt  
\begin{center} \hskip -45pt 
\rotatebox{90}{\hskip 20pt Deflection angle    
~$\log_{10}[\theta^\odot_{\tt gr}], ~[\mu{\rm as}]$} \hskip -45pt
\begin{minipage}[b]{.46\linewidth}
\vskip -25pt
\centering\psfig{figure=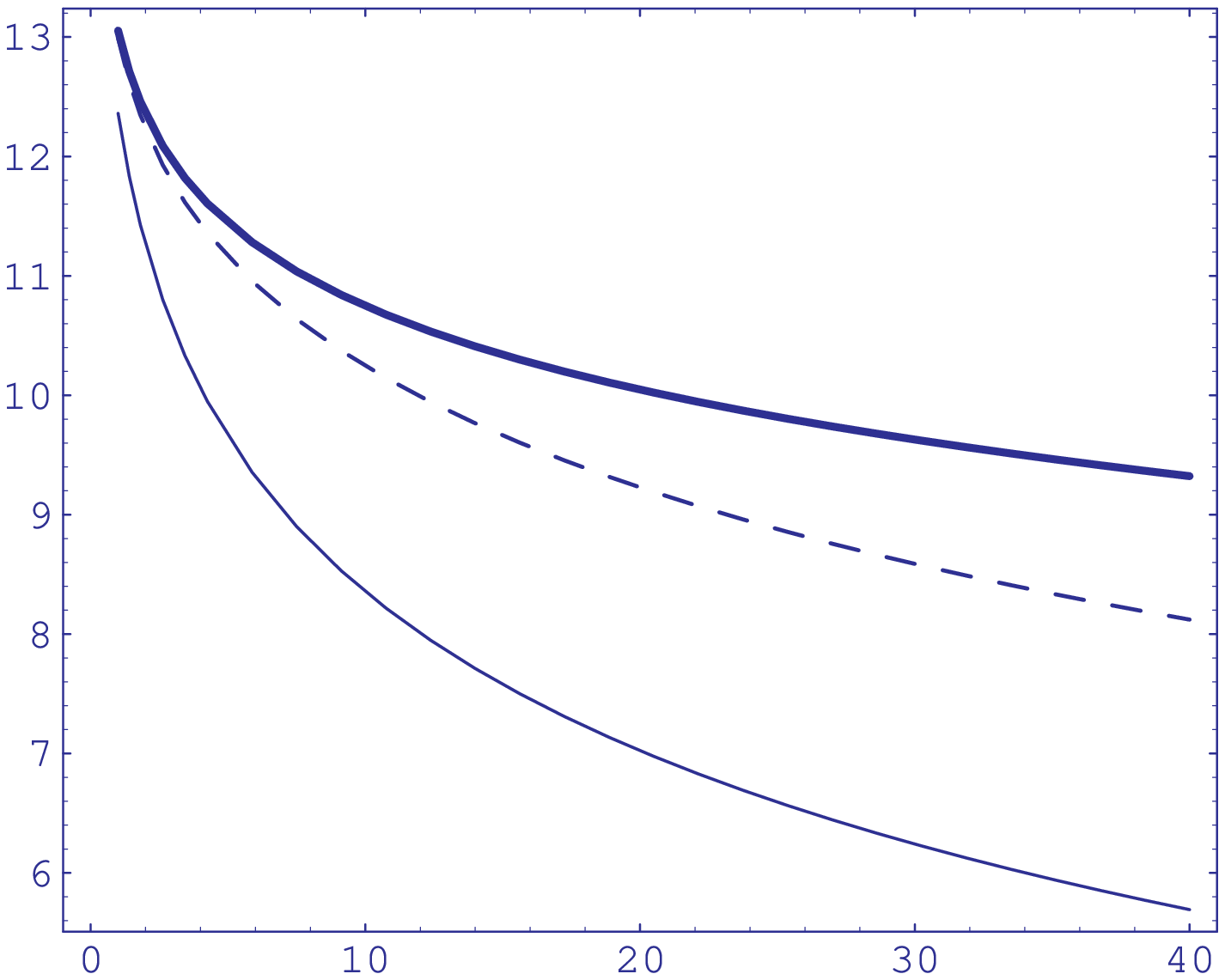,width=83mm,height=63mm}
\rotatebox{0}{\hskip 55pt  Sun-source separation angle,
~$\chi_{1\odot}$~[deg]}
\end{minipage}
\hskip 40pt
\rotatebox{90}{\hskip 20pt  
Deflection angle  ~$\log_{10}[\theta^\odot_{\tt gr}], ~[\mu{\rm as}]$}
\hskip -45pt
\begin{minipage}[b]{.46\linewidth}
\vskip -25pt
\centering \psfig{figure=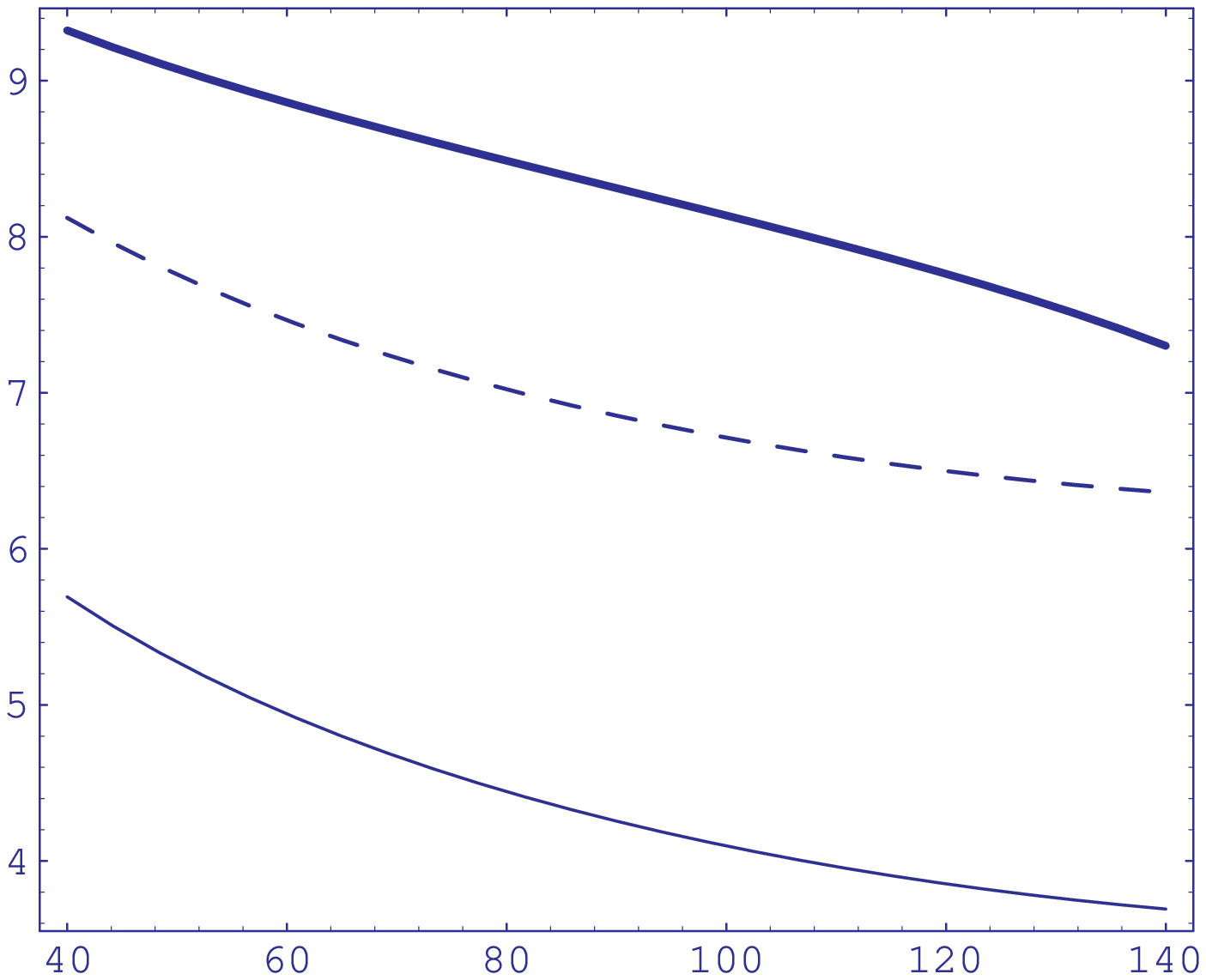,width=83mm,height=63mm}
\rotatebox{0}{\hskip 55pt  Sun-source separation angle,
~$\chi_{1\odot}$~[deg]}
\end{minipage}
     \caption{Solar gravitational deflection of light. On all plots: the upper
thick line is for the  absolute astrometric measurements, 
while the other two are for the differential astrometry. Thus, the  dashed 
line is for the observations over field of regard  of 
FoR$\,\,= 15^\circ$, the lower thick line is for FoR $ = 1^\circ$. 
      \label{fig:solar_defl}}
 \end{center}
\end{figure}

A qualitative presentation of the solar gravitational deflection is given
in the Figure \ref{fig:solar_defl}. The upper thick
line on both plots represents the absolute astrometric measurements, 
while the other two are for the differential astrometry. Thus, the middle dashed 
line is for the observations over the maximal field of regard  of the
instrument FoR$\,\,= 15^\circ$, the lower thick line is for FoR $ =
1^\circ$. 

%
\subsubsection{Second Order Post-Newtonian 
Effects in the Solar Deflection}

One may also want to  account for the
post-post-Newtonian (post-PN) terms (e.g.  $\propto G^2$) as well as the
contributions due to  other PPN parameters (refer to \cite{Will93}).  Thus,   in
the weak gravity field approximation the total deflection  angle
$\theta_{\tt gr}$ has an additional  contribution due the 
post-post-Newtonian terms in the metric tensor.  For the crude estimation
purposes this effect  could be given by the following expression:    
{}
\begin{equation}
\delta\theta_{\tt post-PN}=
\frac{1}{4}(\gamma+1)^2  
\left(\frac{2GM}{c^2d}\right)^2
\left(\frac{15\pi}{16}-1\right)
\left(\frac{1+\cos\chi}{2}\right)^2.
\label{eq:postPN}
\end{equation}

\noindent However, a quick look on the magnitudes of these terms  for the
solar system's bodies suggested that SIM astrometric data will be 
insensitive to the post-PN effects. The   post-PN effects   due to the  Sun
are the largest among those in the solar system. However, even for the
absolute astrometry with the Sun-grazing rays the post-PN terms were 
estimated to be of order $\delta\theta_{\tt post-PN}^\odot= 7~\mu$as 
(see \cite{Turyshev98}). Note that the SIM solar avoidance angle
(SAA) is constraining  the Sun-source separation angle  as $\chi_{1\odot}
\ge$ 45$^\circ$. The post-PN effect is inversely proportional to the square
of the impact parameter, thus reducing the effect to  
$\delta\theta^\odot_{\tt post-PN}\le 3.1 $ nanoarcseconds  on the rim of
SAA. This is why the post-PN effects will not  be  accessible with  SIM.
 
\vskip 20pt
\subsection{Gravitational Deflection of Light by the Jupiter} 

One may obtain the expression, similar to Eq.(\ref{eq:sun_defl}) for
the relativistic deflection of light by the Jovian gravitational
monopole in the following form:    {} 
\begin{equation}
\theta^J_{\tt gr}= (\gamma+1)~\frac{G}{c^2}
\frac{M_J}{r_J}~\frac{1+\cos\chi_{1J}}{\sin\chi_{1J}}=
0.924944\cdot\frac{1+\cos\chi_{1J}}{\sin\chi_{1J}}
~~~~\mu{\sf as},
\label{eq:deflec_jup}
\end{equation}
\noindent with $\chi_{1J}$ being the Jupiter-source
separation angle as seen by the interferometer at the distance $r_J$ from
the Jupiter. For the differential  observations one will have
expression, similar to that  
Eq.(\ref{eq:sun_def_defl}) for the Sun:   
{} 
\begin{equation}
\delta\theta^J_{\tt gr}=(\gamma+1)\frac{G}{c^2} ~
\frac{M_J}{r_J}\frac{\sin\frac{1}{2}(\chi_{2J}-\chi_{1J})}
{\sin\frac{1}{2}\chi_{1J}\cdot\sin\frac{1}{2}\chi_{2J}}=0.924944
\cdot\frac{\sin\frac{1}{2}(\chi_{2J}-\chi_{1J})}
{\sin\frac{1}{2}\chi_{1J}\cdot\sin\frac{1}{2}\chi_{2J}}~~~~\mu{\sf as}, 
\label{eq:deflec_diff_jup}
\end{equation}
  
\noindent where again $\chi_{1J}, \chi_{2J}$ are the Jupiter-source
separation angles for the primary and secondary stars
correspondingly, $\chi_{2J}=\chi_{1J}+\frac{\pi}{12}$
(and $\chi_{2J}=\chi_{1J}+\frac{\pi}{180}$ for the narrow angle
astrometry).   The largest effect will come when SIM and the Jupiter
are at the closest distance from each other   $\sim 4.2 ~$AU.
The Jupiter's angular dimensions from the Earth' orbit for this
situation are  calculated to be
${\cal R}_J=23.24$ ~arcsec,  which correspond to a deflection angle of 
16.419 mas.  Results for some $\chi_{1J}$ are given in the  Table
\ref{tjes}. Note that for the light rays coming perpendicular to the ecliptic 
plane the Jovian deflection will be in the range:
$\delta\alpha_{1J}\sim(0.7\--1.0)~\mu$as!

\begin{figure}[h]
\noindent  
\begin{center} \hskip -45pt 
\rotatebox{90}{\hskip 20pt Deflection angle    
~$\log_{10}[\theta^J_{\tt gr}], ~[\mu{\rm as}]$}
\hskip -45pt
\begin{minipage}[b]{.46\linewidth}
\vskip -25pt
\centering\psfig{figure=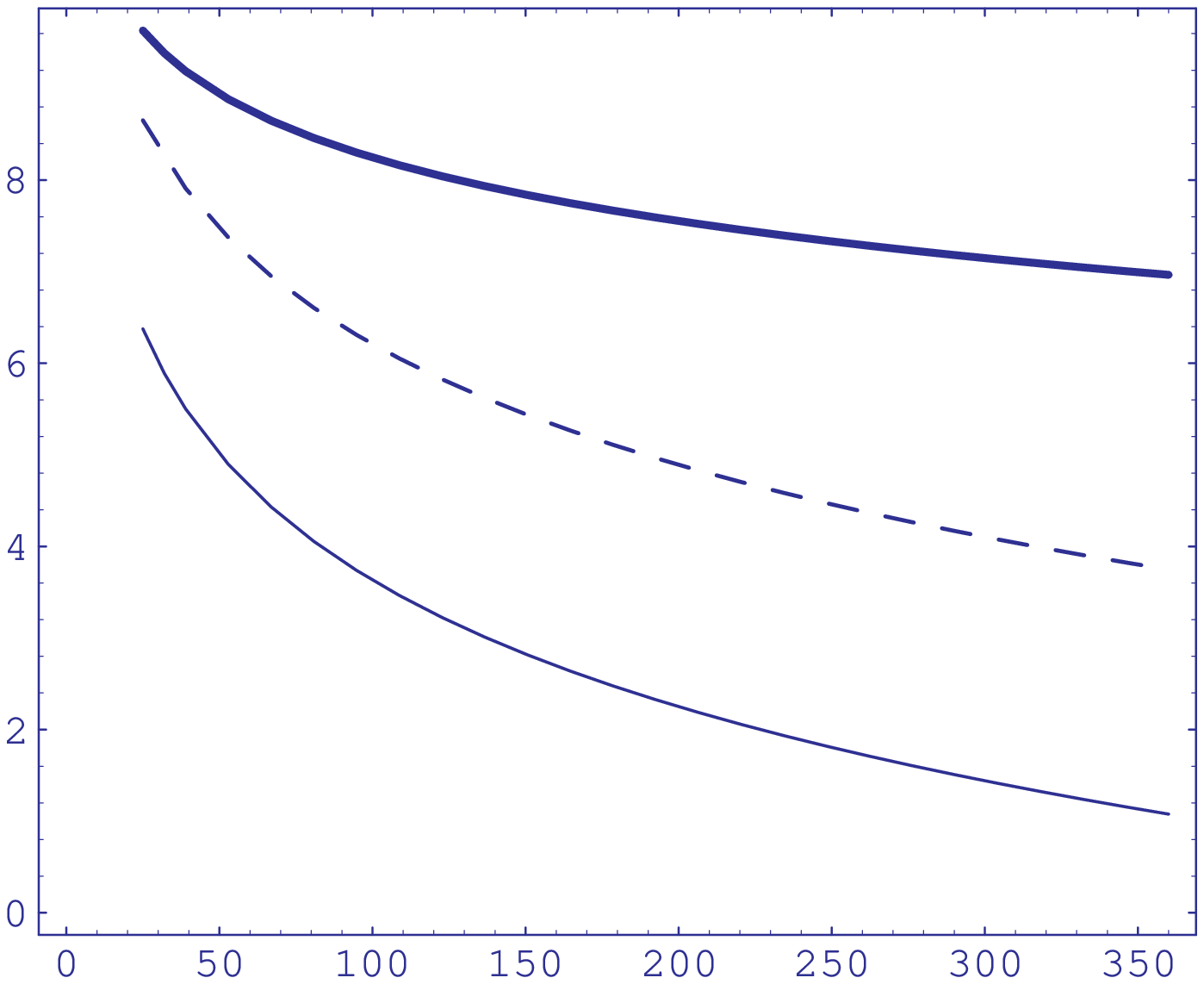,width=85mm,height=70mm}
\rotatebox{0}{\hskip 43pt  Jupiter-source separation angle,
~$\chi_{1J}$~[arcsec]}
\end{minipage}
\hskip 40pt
\rotatebox{90}{\hskip 20pt  
Deflection angle  ~$\log_{10}[\theta^J_{\tt gr}], ~[\mu{\rm as}]$}
\hskip -45pt
\begin{minipage}[b]{.46\linewidth}
\vskip -25pt
\centering \psfig{figure=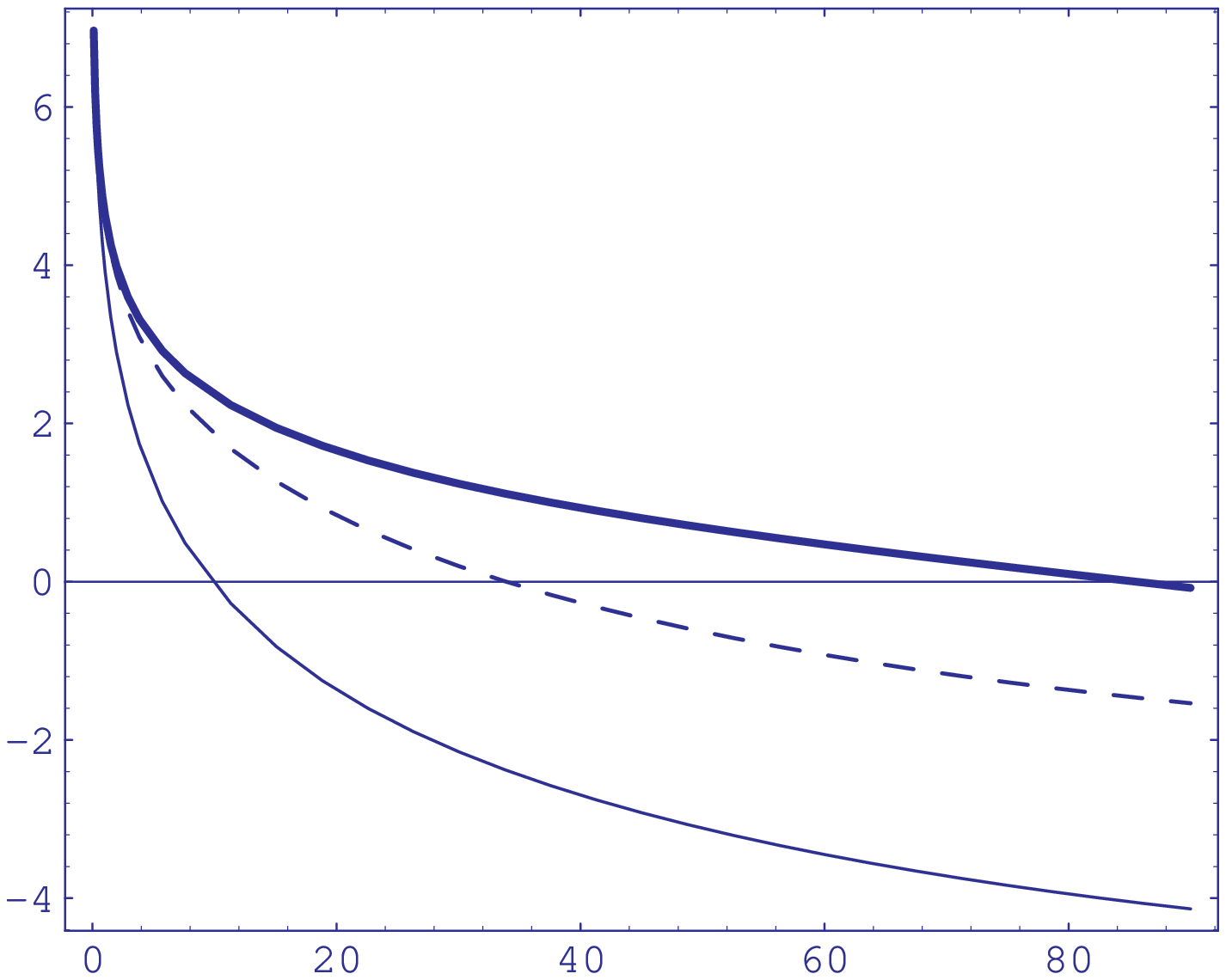,width=85mm,height=70mm}
\rotatebox{0}{\hskip 50pt  Jupiter-source separation angle,
~$\chi_{1J}$~[deg]}
\end{minipage}
     \caption{Jovian gravitational deflection of light. 
      \label{fig:jupiter_defl}}
 \end{center}
\end{figure}

A qualitative behavior of the effect of the gravitational
deflection of light by the Jovian gravity field is plotted in the 
Figure \ref{fig:jupiter_defl}. As in the case of the solar deflection, the upper
thick line on both plots represents the absolute astrometric measurements, 
while the other two are for the differential astrometry (the  
dashed  line is for the observations over FoR$\,\,= 15^\circ$ and 
the lower thick line is for FoR $ = 1^\circ$). 

\vskip 10pt
\begin{table}[h]
\begin{center}
\begin{tabular}{|r|c|c|c|c|c|c|c|c|}
\hline
\tt Jovian   & \multicolumn{8}{c|}{ Jupiter-source
separation angles $\chi_{1J}, ~~$arcsec}  \\
\cline{2-9} 
\tt deflection 
&$23.24''$ &$26''$&$30''$&$60''$&$120''$&$180''$&$360''$
&$90^\circ$\\\hline
$~\theta^J_{\tt gr},~$mas &
16.419 &
14.676 & 
12.719 &
6.360 &
3.180 & 
2.120 &
1.060 &
$0.9~\mu$as\\ \hline
$\delta\theta^J_{\tt gr}[15^\circ],~$mas &
16.412 &
14.669 & 
12.712 &
6.352 &
3.173 & 
2.113 &
1.053 &
$0.2~\mu$as\\ \hline
$\delta\theta^J_{\tt gr}[1^\circ],~$mas &
16.313 &
14.570 & 
12.614 &
6.255 &
3.077 & 
2.019 &
0.964 &
$0.0~\mu$as\\ \hline
\end{tabular} 
\caption{Jovian gravitational monopole deflection  vs. the 
Jupiter-source sky separation angle  $\chi_{1J}$.}
\label{tjes}
\end{center}
\end{table}

\subsection{Gravitational Deflection of Light by the Earth} 

The deflection of light rays by the Earth's gravity field may also be of
interest. The expressions, describing  the relativistic deflection
of light by the Earth' gravitational monopole are given below:   
{} 
\begin{equation}
\theta^\oplus_{\tt gr}= (\gamma+1)~\frac{G}{c^2}
\frac{M_\oplus}{r_\oplus}~\frac{1+\cos\chi_{1\oplus}}{\sin\chi_{1\oplus}}=
0.2446\cdot\frac{1+\cos\chi_{1\oplus}}{\sin\chi_{1\oplus}}
~~~~\mu{\sf as},
\label{eq:deflec_earth}
\end{equation}
\noindent with $\chi_{1\oplus}$ being the Earth-source
separation angle as seen by the interferometer at the distance $r_\odot$ from
the Earth. Relation for the differential astrometric measurements was
obtained in the form:  
 
\begin{equation}
\delta\theta^\oplus_{\tt gr}=(\gamma+1)\frac{G}{c^2} ~
\frac{M_\oplus}{r_\oplus}
\frac{\sin\frac{1}{2}(\chi_{2\oplus}-\chi_{1\oplus})}
{\sin\frac{1}{2}\chi_{1\oplus}\cdot\sin\frac{1}{2}\chi_{2\oplus}}
=0.2446\cdot\frac{\sin\frac{1}{2}(\chi_{2\oplus}-\chi_{1\oplus})}
{\sin\frac{1}{2}\chi_{1\oplus}\cdot\sin\frac{1}{2}\chi_{2\oplus}}
~~~~\mu{\sf as}, 
\label{eq:deflec_diff_earth}
\end{equation}
  
\noindent where, as before, $\chi_{1\oplus}, \chi_{2\oplus}$ are the
Earth-source separation angles for the primary and secondary stars
correspondingly, $\chi_{2\oplus}=\chi_{1\oplus}+\frac{\pi}{12}$
(and $\chi_{2\oplus}=\chi_{1\oplus}+\frac{\pi}{180}$ for the narrow
angle astrometry).   The largest effect will come when SIM and the
Earth are at the closest distance, say at the end of the first 
half of the  first year of the mission,
$r_\oplus=0.05$ AU. The Earth's angular dimensions being measured from 
the spacecraft from that distance are  calculated to be 
${\cal R}^{\tt SIM}_\oplus=175.88401$ arcsec, 
which correspond to a deflection angle of ~ 573.75 $\mu$as. 
The deflection angles   for a few $\chi_{1\oplus}$ are given 
in the Table \ref{tab:sim_hip}.


\begin{table}[h]
\begin{center}
\caption{Solar relativistic deflection angle as a function of the
Earth-source  separation angle $\chi_{1\oplus}$. Results for SIM  are
given for the first half of the first year mission, when the distance
between the spacecraft and the Earth is $\sim$ 0.05 AU.
\label{tab:sim_hip}}\vskip 10pt
\begin{tabular}{|r|c|c|c|c|c|c|c|}
\hline

\tt SIM    & \multicolumn{7}{c|}{$\chi^{\tt SIM}_{1\oplus},$ ~arcsec}   \\
\cline{2-8}
\tt mission & 175.88 & 200 & 360 &
$1^\circ$ & $5^\circ$ & $10^\circ$ & $15^\circ$\\\hline 
$~~\theta_{1\oplus},~\mu$as &
573.8 &
504.7 & 
280.3 &
28.0 &
5.6 & 
2.8 &
1.9 \\ \hline
$\delta\theta_{1\oplus}[15^\circ],~\mu$as &
571.9 &
502.7 & 
278.5 &
26.3 &
4.2 & 
1.7 &
1.0 \\ \hline
$\delta\theta_{1\oplus}[1^\circ],~\mu$as &
547.0 &
478.0 & 
254.8 &
14.0 &
0.9 & 
0.3 &
0.1 \\ \hline
\end{tabular} 

\end{center}
\end{table}

\begin{figure}[h]
\noindent  
\begin{center} \hskip -45pt 
\rotatebox{90}{\hskip 20pt Deflection angle    
~$\log_{10}[\theta^\oplus_{\tt gr}],~[\mu{\rm as}]$}
\hskip -45pt
\begin{minipage}[b]{.46\linewidth}
\rotatebox{0}{\hskip 52pt At the distance of 0.05 AU from the Earth}
\vskip -30pt
\centering\psfig{figure=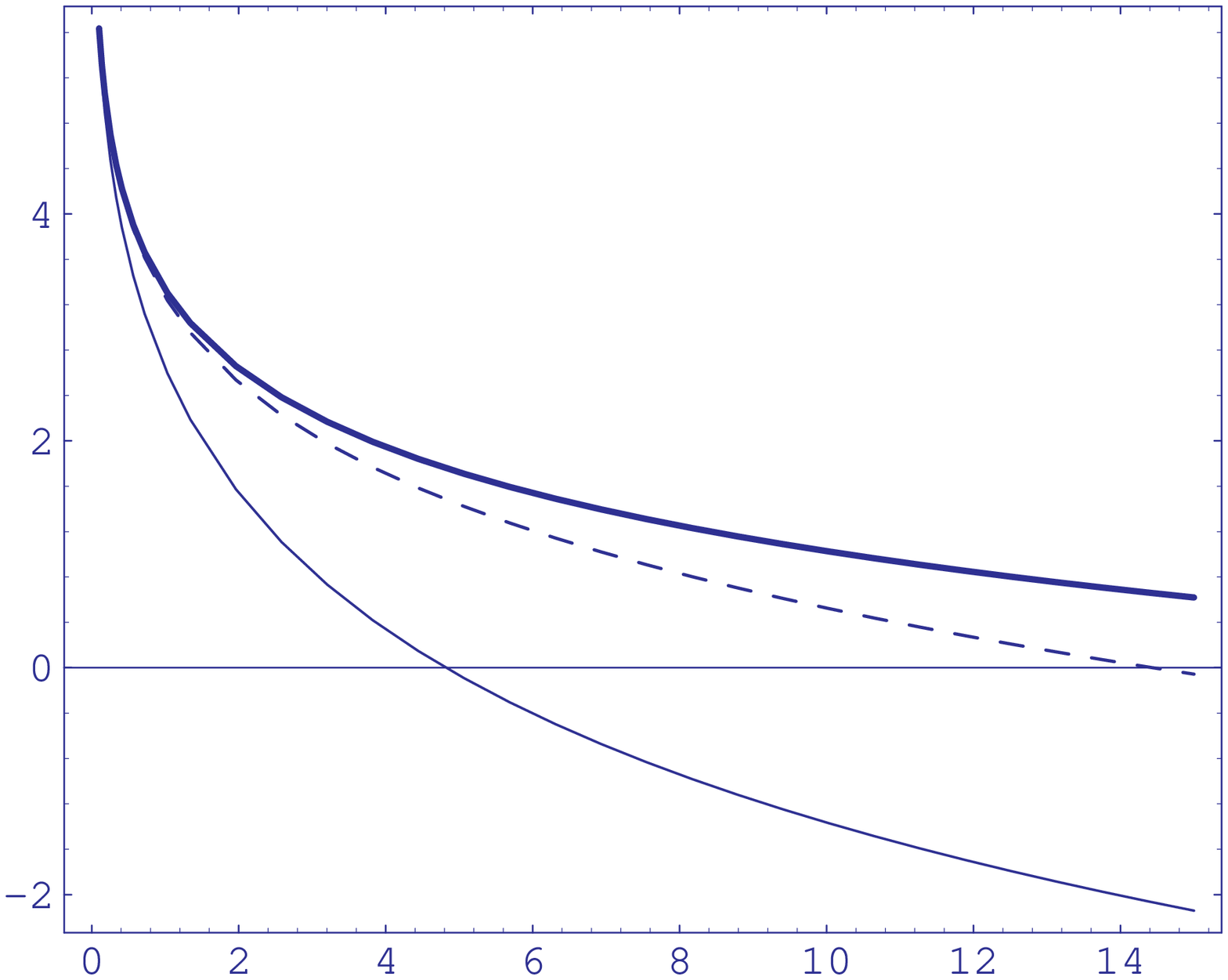,width=85mm,height=70mm}
\rotatebox{0}{\hskip 57pt  Earth-source separation angle,~$\chi_{1\oplus}$~[deg]}
\end{minipage}
\hskip 40pt
\rotatebox{90}{\hskip 20pt  
Deflection angle  ~$\log_{10}[\theta^\oplus_{\tt gr}], ~[\mu{\rm as}]$}
\hskip -45pt
\begin{minipage}[b]{.46\linewidth}
\rotatebox{0}{\hskip 55pt   At the distance of 0.5 AU from the Earth}
\vskip -30pt
\centering \psfig{figure=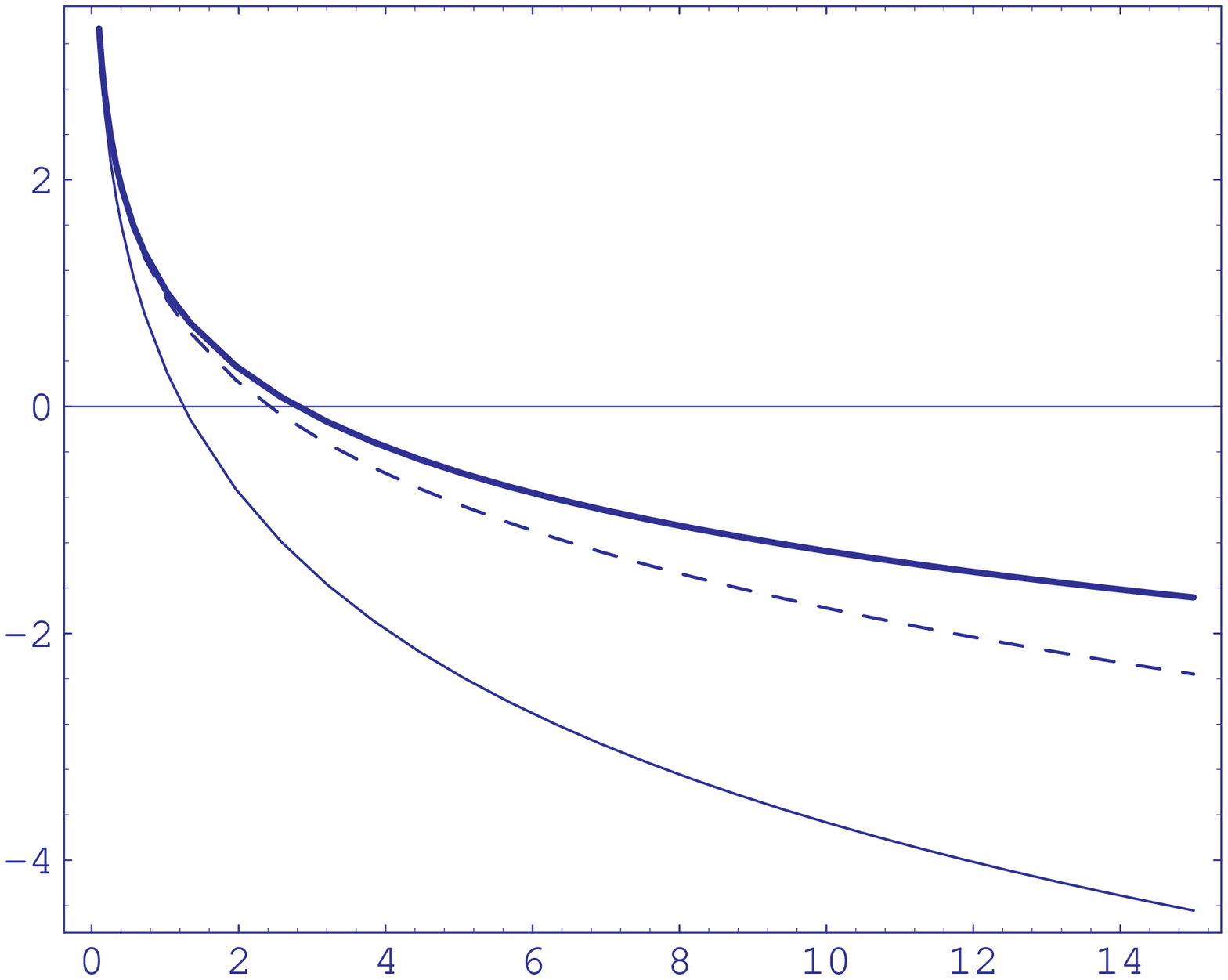,width=85mm,height=70mm}
\rotatebox{0}{\hskip 57pt  Earth-source separation angle,~$\chi_{1\oplus}$~[deg]}
\end{minipage}
     \caption{Gravitational deflection of light in the proximity of the Earth. 
      \label{fig:earth_delf}}
 \end{center}
\end{figure}

In the Figure \ref{fig:earth_delf} we have presented the expected variation
in the magnitude of the Earth' gravity influence as mission progresses.
Thus, the left plot is for the end of the first half of the year of the mission,
when the spacecraft is at the distance of 0.05 AU from the Earth. The  
plot on the right side is for the end of the 5-th year of the mission, when SIM
is at 0.5 AU from Earth.

\section{Constraints Derived From the Monopole Deflection of Light}
\label{sec:constr}

While analyzing the solar gravity field's influence on the future
astrometric observations with SIM, we found several interesting
situations, that may potentially put an additional navigational
requirements. In this section we will consider these situations in a
more detailed way.

\subsection{Accuracy of Impact Parameter and Planetary
Barycentric Position}

To carry out an adequate reduction of observations  with a  
$\Delta \theta_0=\Delta k ~\mu{\rm as}$
accuracy, it is necessary to  determine precisely the value 
of impact parameter of photon's trajectory with respect to 
the body that deflects the photon's motion from the rectilinear one. 
As before, we will present two types of necessary expressions, 
namely for absolute and  differential observations.
By using the equation (\ref{eq:b_defl}), one may present the
uncertainty  $\Delta d_B$ in determining the impact parameter for a
single ray as follows   
{}
\begin{equation}
\Delta d_B= \Delta \theta_0\,
\,\frac{r^2_B\sin^2\chi_{1B}}{2\mu_B}\cdot
 \frac{\cos\chi_{1B}}{1+\cos\chi_{1B}}.
\label{eq:impact_parameter}
\end{equation}
\noindent The corresponding result for differential observations may be
obtained with the help of Eq.(\ref{eq:diff_deflec}) as:
{}
\begin{equation}
\Delta d^{\tt diff}_B= \Delta \theta_0\,
\,\frac{r^2_B\sin^2\chi_{1B}}{4\mu_B}\cdot
\Big[1+\tan\frac{\chi_{1B}}{2}\cdot
\cot\frac{1}{2}(\chi_{2B}-\chi_{1B})\Big].
\label{eq:impact_parameter_diff}
\end{equation}
\noindent Similarly, the uncertainty 
in determining the barycentric distance $r_B$ should be less then
given by the formula  below:
{}
\begin{equation}
\Delta r_B= \Delta \theta_0 \,\frac{r^2_B}{2\mu_B}\cdot
 \frac{\sin\chi_{1B}}{1+\cos\chi_{1B}}.
\label{eq:bar_dist}
\end{equation}
A similar expression for the uncertainty in barycentric
position  $\Delta r^{\tt diff}_B$  for differential observations 
does not produce any  constraints significantly different from those
derived from  Eq.(\ref{eq:bar_dist}), thus we decided not to use it in
our analysis.
Looking at the results presented in the Table \ref{tab6}, one may see
that for astrometric accuracy  $\Delta \theta_0=1~\mu$as our estimates
resulted in fact that one must know the impact
parameters with respect to the center of mass of the Sun with the 
accuracy of $\sim 0.4$ ~km (grazing rays), the Jupiter with the
accuracy of $\sim 4 ~$km and other  big planets with the accuracy of
about  10 ~km. The corresponding estimates are given in the Table
\ref{tab6}.  In order to compare these derived requirements on of the
barycentric positions of the solar system's bodies  with
the current state-of-the-art in their determination, we presented the
best known accuracies in the  Table \ref{tab_best}. The best known
accuracies were taken form DE405/LE405. 
 
\begin{table}
\begin{center}
\caption{Required accuracy of   
barycentric positions and impact parameters for astrometric
observations with accuracy of 1 $\mu$as. The Earth is taken at the distance of
0.05 AU from the spacecraft. Accuracy for the Moon's position  is
given from the geocentric reference frame. } 
\label{tab6}\vskip 10pt

\begin{tabular}{|r|c|c|c||c|c|c|c|} \hline

Solar & \multicolumn{3}{c||}{Required knowledge: grazing rays}  
& \multicolumn{4}{c|}{Required knowledge: differential astrometry}
\\\cline{2-8}
system's & Distance, & \multicolumn{2}{c||}{Impact parameter} & 
\multicolumn{2}{c|}{Impact param. [15$^\circ$]}  &  
\multicolumn{2}{c|}{Impact param. [1$^\circ$]} \\\cline{3-8} 
object & $\sigma_{r_B}, $~km & $\sigma_{d_B}, $~km &
$\sigma_{d_B}, $~mas & $\sigma_{r_B}$,~km & $\sigma_{r_B}$,~mas&
$\sigma_{r_B}$,~km & $\sigma_{r_B}$,~mas \\[2pt]\hline\hline

Sun     & 85.45 & 0.39& 0.55 &  0.40 & 0.55  & 0.50 & 0.69 \\

Sun   at 45$^\circ$  & 1.5$\times10^4$ & 7.6$\times10^3$
        & 10$''$.49 & 3.81$\times10^4$  
        & 52$''$.53 & 4.45$\times10^5$ & 613$''$.66\\ 

Moon    & 2.8$\times10^5$ & 67.14 &  1$''$.85 
        & 67.17  & 1$''$.85  &68.00 & 1$''$.88 \\

Mercury &1.1$\times10^6$& 29.39 & 66.16  & 29.41 
        & 66.16 & 29.45 & 66.25\\

Venus   & 8.4$\times10^4$ & 12.18 & 61.00  & 12.28 & 61.20  
        & 12.38 & 62.70 \\

Earth-Moon  & 1.3$\times10^4$ & 11.14 & 306.55  
            & 11.15 & 307.47 & 11.66 & 321.54\\

Mars    & 6.8$\times10^5$ & 29.29 & 77.11 & 29.30  & 77.14  
        & 29.37 & 77.33\\ 

Jupiter & 3.8$\times10^4$ & 4.31 & 1.42 & 4.32 & 1.42& 4.34 & 1.42\\

Jupiter at 30$''$ & 4.9$\times10^4$ & 7.14 
        & 2.34 & 7.20 & 2.36 & 7.25 & 2.38\\ 

Saturn  & 2.2$\times10^5$ & 10.32 & 1.66 & 10.34   
        & 1.66 & 10.36 & 1.66\\

Uranus  & 1.2$\times10^6$ & 11.27 & 0.86 & 11.28   
        & 0.86 & 11.29 & 0.86\\

Neptune & 1.7$\times10^6 $& 10.04 & 0.47 & 10.04  
        & 0.47 & 10.04 & 0.47 \\

Pluto    & 2.0$\times10^9$ & 1133.92 & 40.7
         & 1133.93   & 40.7 & 1133.96& 40.7\\\hline

\end{tabular} 
\end{center} 
\end{table}


\vskip 60pt 
\begin{table}
\begin{center}
\caption{The best known accuracies of   
barycentric positions and masses for the solar system's objects
 derived form DE405/LE405. Planetary masses taken from  
\cite{Yoder95}.} \label{tab_best}
\vskip 10pt 
\begin{tabular}{|r|c|c|c||c|} \hline

Solar   
& \multicolumn{3}{c||}{Knowledge of barycentric position}
& Knowledge of  \\\cline{2-4}
system's &  \multicolumn{2}{c|}{Best known,} & Method used 
& planetary masses,   \\\cline{2-3}  
object & $\sigma_{r_B}, $~km & $\sigma_{r_B}, $~mas & 
for determination & 
$\Delta M_B/M_B$   \\[2pt]\hline\hline

Sun    &  362/725 & $0''.5/1''.0$&  
        Optical meridian transits  &  3.77$\times10^{-10}$   \\


Moon    & 27 ~cm & 7.4 $\mu$as & LLR, 1995 
& 1.02$\times10^{-6}$   \\

Mercury & 1 & 2.25& Radar ranging & 4.13$\times10^{-5}$   \\

Venus   & 1  & 4.98& Radar ranging & 1.23$\times10^{-7}$   \\

Earth   & 1 & 27.58 & Radar ranging  & TBD$\times10^{-6}$   \\

Mars    & 1 & 2.63& Radar ranging & 2.33$\times10^{-6}$   \\ 

Jupiter & 30 & 9.84& Radar ranging & 7.89$\times10^{-7}$  \\


Saturn  & 350 & 56.24& Optical astrometry & 2.64$\times10^{-6}$   \\

Uranus  & 750 & 57.00 & Optical astrometry & 3.97$\times10^{-6}$   \\

Neptune & 3,000 & 141.67& Optical astrometry & 2.19$\times10^{-6}$   \\

Pluto/Charon    & 20,000   & 717.40 & Photographic astrometry 
        & 0.014  \\\hline

\end{tabular} 
\end{center} 
\end{table}


\begin{table}
\begin{center}
\caption{Required accuracy of the planetary masses,  the  PPN parameter
$\gamma$ and the uncertainty in the attitude determination  for
the astrometric error allocation of 1 $\mu$as.} 
\label{tab:masses_gamma}\vskip 10pt
\begin{tabular}{|r||c||c|c||c|} \hline

Solar & Masses, & \multicolumn{2}{c||}{PPN parameter $\gamma$} & 
Attitude accuracy\\\cline{3-4} 
system's object & $\Delta M_B/M_B$ & $\Delta \gamma$ &
$\Delta \gamma_{\tt diff}~[15^\circ]$ & $[15^\circ]$,~
$\Delta(\epsilon-\alpha_B)$  
\\[2pt]\hline\hline

Sun     & 5.7$\times10^{-7}$ & 1.1$\times10^{-6}$ 
        & 1.2$\times10^{-6}$ & ~~~0$''$.124 \\

Sun   at 45$^\circ$  & 1.0$\times10^{-4}$
        & 2.0$\times10^{-4}$ & 7.1$\times10^{-4}$ & 73$''$.23~\\

Moon    & 3.8$\times10^{-2}$ & 7.7$\times10^{-2}$ &7.7$\times10^{-2}$ 
        & ~2$^\circ$.21 \\

Mercury & 1.2$\times10^{-2}$ & 2.4$\times10^{-2}$ 
        & 2.4$\times10^{-2}$& ~0$^\circ$.69\\

Venus   & 2.0$\times10^{-3}$ & 4.1$\times10^{-3}$ 
        & 4.1$\times10^{-3}$ & ~0$^\circ$.12\\

Earth   & 1.7$\times10^{-3}$ & 3.4$\times10^{-3}$  
        & 3.5$\times10^{-3}$ & 361$''$.00~~~\\

Mars    & 8.6$\times10^{-3}$ & 1.7$\times10^{-2}$   
        &1.7$\times10^{-2}$ & ~0$^\circ$.49\\

Jupiter & 6.1$\times10^{-5}$ & 1.2$\times10^{-4}$  
        &1.2$\times10^{-4} $& 12$''$.38  \\

Jupiter at 30$''$ & 7.9$\times10^{-5}$ & 1.5$\times10^{-4}$  
       &1.6$\times10^{-4}$ & 16$''$.50\\

Saturn  & 1.7$\times10^{-4}$ & 3.4$\times10^{-4}$  
        & 3.4$\times10^{-4}$ & 35$''$.07\\

Uranus  & 4.6$\times10^{-4}$ & 9.2$\times10^{-4}$ 
        & 9.2$\times10^{-4}$ & 94$''$.88\\

Neptune & 3.9$\times10^{-4}$ & 7.9$\times10^{-4}$  
        & 8.0$\times10^{-4}$ & 82$''$.51\\

Pluto    & 0.35 & 0.71  &0.71 & 20$^\circ$.34\\\hline

\end{tabular} 
\end{center} 
\end{table}

\subsubsection{Need for Improvement of Knowledge  of 
Planetary Positions}
 
One may see that the present accuracy of knowledge of the inner planets' 
positions  from the Table \ref{tab_best} is given by the radio
observations and it is  even better than the level of relativity
requirements given in  the Table \ref{tab6}. However, the positional
accuracy for the outer planets is  significantly below the required
level. The SIM observation program should include the astrometric
studies  of the outer planets 
in order to minimize the errors in their positional accuracy determination.
Thus, in order to get the radial uncertainty 
in Pluto's ephemeris with accuracy below 1000 ~km, it is necessary only 
4 measurements of Pluto's position, taken sometime within a week of the
stationary points,  spread over 3 years. Each measurement could be taken 
with an accuracy of about $200 ~\mu$as, as suggested by \cite{Standish95}.   
Additionally, one will have to  significantly lean on the 
radio observations in order to 
conduct the reduction of the optical data with an accuracy of a few 
$\mu$as.  For this reason one will have to use  the precise catalog of the
radio-sources and to study the problem of the  radio
and  optical reference frame ties  
(\cite{Standish95,Standish_etal95,Folkner94}).

An important way to improve the accuracy of the  
positions of the outer planets may be offered by the current program of 
the deep space exploration. Thus, one may expect a factor of 3 improvement 
in positional accuracy of the Jupiter and its satellites with the 
completion of  the Galileo mission. The Cassini mission will be in the
Saturn's vicinity  at the time close to the SIM's active
astrometric  campaign | 2009-2015.
A factor of $3\times 10^2$ improvement in the Saturn's system 
positional accuracy may be expected.
The Doppler, range and range-rate measurements to the  spacecraft, 
combined together with the ground-based VLBI  methods will
significantly improve the positional accuracy  for the 
bodies in the solar system.  This will help to increase the
overall accuracy of the SIM astrometric observations via a frame tie
to the radio and the dynamical reference frames. 

From the other hand, the accuracy of a single measurement with SIM is
expected to be of the order of $\sigma_\alpha=8~\mu$as.
If the uncertainty in positions may contribute only to about 10\% of
the total variance, thus 
$\Delta\theta_0=\sqrt{0.1}\sigma_\alpha=2.53~\mu$as. This fact  
relaxes requirements presented in the Table \ref{tab6}. However, even
though the requirements are still much smaller the current best
knowledge, it may be the case when SIM actually will significantly
improve  positions of outer planets of the solar system simply as a
by-product of it's astrometric campaign. To correctly address this
problem one needs to perform a full-blown numerical simulation with a
complete model  for the SIM instrument.

\subsection{Accuracy of Planetary Masses and the PPN Parameter $\gamma$}

The uncertainty 
in determining the solar and the planetary masses $\Delta M_B$ 
should be less then given by the formula below:
{}
\begin{equation}
\frac{\Delta M_B}{M_B}= \Delta \theta_0~\frac{r_B}{2\mu_B}\cdot
 \frac{\sin\chi_{1B}}{1+\cos\chi_{1B}}.
\label{eq:planet_mass}
\end{equation}
The corresponding estimates for $\Delta \theta_0= 1~\mu$as are presented in
the Table \ref{tab:masses_gamma}. Thus, the presently available values
for the planetary masses, given in Table \ref{tab_best}, are more then
sufficient to fulfill the general relativistic  
requirements.  

Similarly, the uncertainty 
in determining the PPN parameter $\gamma$ should be less then
given by the following expression:
{}
\begin{equation}
 \Delta \gamma = \Delta \theta_0~\frac{r_B}{\mu_B}\cdot
 \frac{\sin\chi_{1B}}{1+\cos\chi_{1B}}.
\label{eq:gamma}
\end{equation}
\noindent Finally, the relation for the differential astrometric
measurements one obtains in the form:
\begin{equation}
 \Delta  \gamma_{\tt diff} = \Delta \theta_0~\frac{r_B}{\mu_B}\cdot
\frac{\sin\frac{1}{2}\chi_{1B}\sin\frac{1}{2}\chi_{2B}}
{\sin\frac{1}{2}(\chi_{2B}-\chi_{1B})}.
\label{eq:gamma_diff}
\end{equation}

\noindent The corresponding results for uncertainties in the planetary
masses and PPN parameters $\gamma$ needed for $\Delta\theta_0= 1~\mu$as
astrometric accuracy  are given in the  Table \ref{tab:masses_gamma}. 
Presently the best known determination of the PPN parameter $\gamma$ is
$|\gamma-1|\le3\times 10^{-4}$ and  was  given by \cite{Eubanks97}.  [Note
that the authors have made a very  first attempt to include  the post-PN
effects (e.g $\propto G^2$)  into their model and  corresponding VLBI data
analysis.] Thus, the
value of this PPN parameter  will have to be improved either before SIM
will be launched or by the mission itself.

\subsection{Astrometric Test of General Relativity}

\subsubsection{Solar Gravity Field as a Deflector}

To model the astrometric data to the nominal measurement accuracy  will
require including the effect of general relativity on the propagation  of light.  
In the PPN framework, the parameter
$\gamma$ would be part of  this model and could be estimated in
global solutions. The astrometric residuals may be tested for any
discrepancies with the prescriptions of general relativity. 
To address this problem in a more detailed way, one will have to use the 
astrometric model for the instrument including the
information about it's position in the solar system, it's attitude
orientation in the proper reference frame, the time history of different
pointings and their durations, etc.   This information then should be
folded into the parameter estimation program that will use a model  based
on the expression,  similar to that given by  
Eq.(\ref{eq:defl_chi_diff}). In addition, due to the geometric
constraints of the spacecraft's orbit in the solar system, one may
expect that  solution for  the parameter
$\gamma$  will be highly correlated with the solution for parallaxes.  

Taking into account the fact that presently we are lacking the existence
of a real data, we may only estimate a possibility of increasing the
accuracy of the parameter's $\gamma$ determination.  Thus, the
estimates  from the Table
\ref{tab:mon} have demonstrated  that   effect of gravitational
deflection of light may be used to estimate the value  of PPN parameter
$\gamma$ at a scientifically important level. Most important is that the
corresponding result could be obtained simply as a by-product  of the
SIM astrometric campaign (see 
\cite{Turyshev98}). For the crude estimation purposes
one may present the expected accuracy of  the parameter $\gamma$
determination in a single astrometric measurement as: 
{}
\begin{equation}
\Delta\gamma   = \Delta  \theta_0  ~
\frac{r^{\tt SIM}_\odot}{\mu_\odot}~
\frac{\sin\frac{1}{2}\chi_{1\odot}\,\sin\frac{1}{2}\chi_{2\odot}}
{\sin\frac{1}{2}(\chi_{2\odot}-\chi_{1\odot})},
\label{eq:gamma_sun}
\end{equation}
\noindent where $\Delta\theta_0$ is the largest
tolerable error in the total error budget allowed for the stellar
aberration due to relativistic deflection  of light in the solar system. 
   

The relativity test will be enhanced by scheduling measurements  of stars
as close to the Sun  as possible. Despite the fact that  during
it's observing campaign, SIM will never be  closer to the Sun than
$45^\circ$,  it  is still will allow for an  accurate determination of
this PPN parameter. Thus, a single astrometric measurement with SIM is
expected to be with an accuracy of $\sigma_\alpha=8~\mu$as. 
It seems to be a  reasonable assumption that a contribution of any
component of the total error budget should not exceed  
10\% of the total variance  a single accuracy of $\sigma^2_\alpha$. This
allows to estimate the correction factor $\Delta \theta_0$ in
Eq.(\ref{eq:gamma_sun})  to be
$\Delta \theta_0=8\sqrt{0.1}=2.52982~\mu$as.    Thus, at the rim of the
solar avoidance angle,  
$\chi_{1\odot}=45^\circ$,  one  could  determine  this parameter with an
accuracy  $|\gamma-1|\sim   1.79\times 10^{-3}$ in a single
measurement. 
When the mission progresses the accuracy of this experiment will improve
as $1/\sqrt{N}$, where $N$ is the number of independent observations. 
With $N \sim 5000$, SIM may achieve accuracy of $\sigma_\gamma \sim  
2.4\times 10^{-5}$  in   astrometric tests of general relativity  in the
solar gravity field.

\subsubsection{GR Test in the Jovian and Earth' Gravity Fields}

It is worth noting that  one could perform  relativity experiment not
only with the Sun, but also with the Jupiter and the Earth. 
In fact,  for the proposed SIM's observing mode, the accuracy of
determining of the parameter $\gamma$ may be  even better than that
achievable with the Sun. Indeed,  with the same assumptions as above, 
one may achieve a single measurement the accuracy    
of $|\gamma-1|\sim 4.0 \times 10^{-4}$ determined via deflection of light
by Jupiter. However, astrometric observations in the  Jupiter's vicinity
are the  targeted observations. One will have to specifically plan those
experiments in advance. This fact is minimizing the number of possible
independent observations and, as a result, the PPN parameter $\gamma$
may be obtained with accuracy  of about
$\sigma_\gamma\sim 1.3 \times 10^{-5}$ with astrometric experiments  in
the Jupiter's gravity field (note that only $N \sim $ 1000 needed).  
For a long observing times $\sim 10^3~$sec  the Jupiter's orbital motion
could significantly contribute to this experiment
(see Sec.\ref{sec:orb_motion} for details).  
 
Lastly, let us mention that the  experiments conducted in the Earth's gravity
field,  could also determine this parameter to an accuracy 
$|\gamma-1|\sim8.9\times 10^{-3}$ in a single measurement (which in
return extends  the measurement of the gravitational bending of light
to a different mass and distances scale, as shown by  \cite{Gould93}). One may
expect a large statistics gained from both the astrometric observations
and the telecommunications with the spacecraft. This, in return,  will 
significantly enhance the overall solution for
$\gamma$  obtained in the Earth' gravitational environment.  
 
\subsection{Baseline Orientation and the Attitude Control Accuracy}

At this point we would like to study the effect of the attitude 
determination uncertainty on the accuracy of the astrometric data
correction for the effect of gravitational deflection of light. In order to  
estimate  the tolerable uncertainty in determining the  orientation of
the baseline vector $\vec{b}$ and the vector of spacecraft's position
with respect to the deflector, $\alpha_B$,  we
will use the equation Eq.(\ref{eq:defl_chi_diff}).    
With the help of this equation one obtains the following expression:
{}
\begin{equation}
\Delta(\epsilon-\alpha_B)  = 
\frac{\Delta \theta_0}{\cos(\epsilon-\alpha_B)} ~
\frac{r_B}{2\mu_B}~
\frac{\sin\frac{1}{2}\chi_{1B}\,\sin\frac{1}{2}\chi_{2B}}
{\sin\frac{1}{2}(\chi_{2B}-\chi_{1B})}.
\label{eq:attitude}
\end{equation}
Thus,  the case with $~|\cos(\epsilon-\alpha_B)|=1$ is the most accuracy
demanding orientation of the vectors involved. Corresponding estimates for
$\Delta \theta_0= 1~\mu$as are presented in the Table \ref{tab:masses_gamma}.
Deflection of light by the Jupiter puts the most stringent requirement on
the attitude control and baseline orientation of
$\Delta\epsilon\approx\Delta \alpha_B =$ 16$''$.50/$\sqrt{2}=11''$.67. 
However,  in accord to the current error budget allocations, which is
bookkeeping  a much smaller number (e.g. $\sim$ few mas), this requirement will
be easily met.

\subsection{Stellar Aberration Introduced by the Orbital 
Motion of a Deflector}
\label{sec:orb_motion} 

The orbital motion of the deflecting bodies  could significantly 
contribute to the relativistic deflection measurement.  To
estimate this influence, let us assume that  during the experiment the
deflecting body moves with   velocity
$\vec{v}_B$. This motion will result in the time-dependent change 
of the impact parameter $d_B$. Such a variation 
could  produce an additional  angular drift with the rate 
$\dot{\theta}^B_{\tt gr}$, in addition to the  static monopole 
deflection Eq.(\ref{eq:b_defl}).  For the estimation purposes, it is 
convenient to express the total effect of the 
monopole deflection Eq.(\ref{eq:b_defl}) and this aberrational
correction  in terms of the deflector-source  sky separation angle
$\chi_{B_0}$ at the beginning of the experiment.    

Thus, a circular orbital motion of the  deflecting body produces a  
drift in the deflector-source separation angle on the sky, 
$\chi_B(t)$, given by the
expression {}
\begin{equation}
\chi_B(t) =\chi_B(t_0)+{\dot\chi}_{B_0}\cdot (t-t_0)+{\cal O}(t^2),
\label{eq:drift}
\end{equation}
\noindent where
$\chi_B(t_0)=\chi_{B_0}$ is the initial observing angle,  and  
$\dot{\chi}_{B_0}={v_B/r_B}$  is the rate of corresponding angular
drift. Assuming that ${\dot\chi}_{B_0}$ is small and for short 
time spans $\Delta t$, one may expand the quantities in the terms of the small
parameter $(v_B\Delta t)/(r_B \chi_{B_0})$.  [This is done  for estimation purposes
only. In a real situation there may not be a small parameter at all. In this
case a full-blown numerical integration should be used instead.] 

\subsubsection{Rate of Absolute Drift Due to Planetary Motion}

As a result of a simplification discussed above, the total
time-dependent effect of the gravitational deflection of light  may be
presented by the  expression  (similar to the concept of a
retarget action): {}
\begin{equation}
\theta^B_{\tt gr}(t)= \theta^B_{\tt gr}(t_0)
-\dot{\theta}^B_{\tt gr}\cdot (t-t_0) +
{\cal O}(\Delta t^2), 
\label{eq:drift_abs}
\end{equation}

\noindent where the first term is the static gravitational deflection
angle at the beginning of the experiment:
\begin{equation}
\theta^B_{\tt gr}(t_0)=
(\gamma+1)~\frac{\mu_B}{r_B}~\frac{1+\cos\chi_{B_0}}
{\sin\chi_{B_0}}  
\label{eq:const_term}
\end{equation}
\noindent with the values for the solar system bodies (for grazing rays!, e.g. 
$r_B\sin\chi_{B_0}=d_B={\cal R}_B$) given in the Table \ref{tab:mon}. 
The second term on the right-hand side of  Eq.(\ref{eq:drift}) is 
the rate of the angular drift due to the planetary motion. This quantity may be
presented in the following form: 
\begin{equation}
\dot{\theta}^B_{\tt gr}= 
\frac{\mu_B~v_B}{r^2_B\,\sin^2\frac{1}{2}\chi_{B_0}}. 
\label{eq:teta_dot}
\end{equation}

Magnitudes of the angular drifts $\dot{\theta}^B_{\tt gr}$
introduced by the orbital motion of the solar system bodies are given in
the  Table \ref{tab:orb_motion}.

\subsubsection{Rate of Differential Drift Introduced by Planetary
Motion} 

The  expressions for the case of differential
observations  may obtained with the help of 
Eq.(\ref{eq:diff_deflec}). We will use the same assumptions on the
smallness of the quantities involved as were used above for the case of
absolute astrometry. As a result, a linear  drift in the planetary
position, Eq.(\ref{eq:drift}), introduces a time-variation in the
gravitational light bending effect,  $\delta\theta^B_{\tt gr}(t)$, as
given by the expressions below: {}
\begin{equation}
\delta\theta^B_{\tt gr}(t)= \delta\theta^B_{\tt gr}(t_0)
-\delta\dot{\theta}^B_{\tt gr}\cdot (t-t_0)+
\dot{\lambda}^{B}_{\tt SIM}\cdot (t-t_0) +
{\cal O}(\Delta t^2), 
\label{eq:drift_diff}
\end{equation}

\noindent where the first term is the static differential 
deflection angle at the beginning of observations
\begin{equation}
\delta\theta^B_{\tt gr}(t_0)=
(\gamma+1)~\frac{\mu_B}{r_B}~
\frac{\sin\frac{1}{2}(\chi_{2B_0}-\chi_{1B_0})}
{\sin\frac{1}{2}\chi_{1B_0}\cdot\sin\frac{1}{2}\chi_{2B_0}} 
\label{eq:const_term_diff}
\end{equation}
\noindent with the values for the solar system bodies presented in the
Table \ref{tab:mon}.  The second term on the right-hand side of 
Eq.(\ref{eq:drift_diff}) is  the rate of the differential angular drift
due to the planetary motion. This quantity is given as follows: 
{}
\begin{equation}
\delta\dot{\theta}^B_{\tt gr}= \frac{\mu_B~v_B}{r^2_B}
\Big[\frac{1}{\,\sin^2\frac{1}{2}\chi_{1B_0}}-
\frac{1}{\,\sin^2\frac{1}{2}\chi_{2B_0}}\Big]. 
\label{eq:theta_dot}
\end{equation}
\noindent The last term in the Eq.(\ref{eq:drift_diff}),
$\dot{\lambda}^{B}_{\tt SIM}$,  is introduced by any temporal drifts in
the accuracy of the SIM instrument during  observations:
{}
\begin{equation}
\dot{\lambda}^B_{\tt SIM}= \frac{\mu_B}{r_B}
\frac{\Delta\dot{\chi_0}}{\,\sin^2\frac{1}{2}\chi_{2B_0}}=
\frac{\mu_B}{r_B}
\frac{\Delta\dot{\chi_0}}{\,\sin^2\frac{1}{2}(\chi_{1B_0}+\Delta\chi_0)}, 
\label{eq:lambda_dot}
\end{equation}
\noindent where $\Delta\chi_0= \chi_{2B_0}- \chi_{1B_0}$ is the 
angular separation between the  two sources at the beginning of the
observation and
$\Delta\dot{\chi_0}=\dot{\chi}_{2B_0}-\dot{\chi_{1B_0}}$ is any time drift
in estimating this  separation introduced by
the instrument [e.g. temporal drifts inside the  observed tile due to 
possible time-varying drifts in the instrument's metrology].  However,
it turns out that  this effect  is not important for our study. Indeed,
even for the  most intense gravitational environment, at the solar
avoidance angle with
$\chi_{1B_0}=45^\circ$, and for the maximal star
separation $\Delta\chi_0=15^\circ$, a constant linear drift with 
the rate of 
$\Delta\dot{\chi_0}=50$ mas/s, produces a total effect of only 
$\dot{\lambda}^B_{\tt SIM}=0.002~\mu$as/sec.
 

\begin{table}[t]
\begin{center}
\caption{Relativistic planetary aberration of light due to their 
barycentric orbital motion. For the purposes of this study, we assumed
SIM at a fixed position in the solar system with $v_{\tt SIM}=0$. 
Aberration due to the solar system's  galactocentric
motion is unobservable.    
\label{tab:orb_motion}}\vskip 10pt
\begin{tabular}{|r|c||c|c|c| } \hline

Solar system's  & Velocity,  & $\dot{\theta}^B_{\tt gr},$ 
&$\delta \dot{\theta}^B_{\tt gr}[15^\circ],$ & 
$\delta \dot{\theta}^B_{\tt gr}[1^\circ],$ \\ object &  km/sec  & $\mu$as/s &
$\mu$as/s & 
$\mu$as/s\\[2pt]\hline\hline

Sun  (galactic)  & 220 & 553.5 & 553.2 & 558.9    \\

 Sun   & 0.013 & 0.033 & 0.033 & 0.031   \\

Sun  at 45$^\circ$   & -same- & 4$\times10^{-7}$ & 5$\times10^{-7}$ &
5$\times10^{-8}$   \\

Moon      & 0.04 & 6$\times10^{-4}$ 
          & 6 $\times 10^{-4}$  & 6$\times10^{-4}$    \\

Mercury  & 47.87 & 1.63 & 1.63 & 1.63    \\

Venus    & 35.05 & 2.86 & 2.86 & 2.86  \\

Earth    & 29.80 & 2.68 & 2.71 & 2.71 \\
 
Mars     &  24.14 & 0.82 & 0.82 & 0.82  \\

Jupiter  & 13.1 & 3.04 & 3.04 & 3.04   \\

Jupiter  at 30$''$ & -same- & 1.82  & 1.82 & 1.82   \\

Saturn   &  9.63 & 0.93 & 0.93 & 0.93   \\

Uranus   & 6.81 & 0.60 & 0.60 & 0.60    \\

Neptune  & 5.44 & 0.54 & 0.54 & 0.54   \\
 
Pluto    & 4.75 & $4\times10^{-3}$ 
         & $4\times10^{-3}$ & $4\times10^{-3}$ \\\hline

\end{tabular} 
\end{center} 
\end{table}

The quantities characterizing the dynamical astrometric environment in
the vicinity of the solar system's bodies are presented in  the Table
\ref{tab:orb_motion}. Due to the fact that the differential effect
behaves as $\sim[1/\sin^2(\tt small_{-}angle)-1/\sin^2(\tt
small_{-}angle+FoR/2)]$ it is almost insensitive to the sizes of the two
available fields of regard. Thus, independently on the size of the available
field of regard,  the consideration of  the orbital velocity of planet's
motion   turns out to be  a significant issue, especially for the Jupiter and
some inner planets. One will have to   account for this effect during a long
exposure   observations, say for $t\sim10^3$~sec.

Concluding, let us mention that the presented estimates were  given for
the  static gravitational field in the barycentric RF. Analysis of a
real experimental situation should  consider  a non-static gravitational
environment of the solar system  and should include 
the description of   light propagation in a different 
RFs involved in the experiment. Additionally, the  
observations will be affected by the   relativistic 
orbital dynamics of the spacecraft.   

\subsection{Solar Acceleration Towards the Galactic Center}

The Sun's absolute velocity with respect to a cosmological
reference frame was measured photometrically:
it shown up as the dipole anisotropy of the cosmic microwave 
background. The Sun's absolute acceleration with respect to a
cosmological reference frame can be measured astrometrically: it
will show up as proper motion of quasars. 

The aberration due to the solar system's 
galactocentric motion will not be observable. However, the rate of this 
aberration will produce an apparent proper motion for the observed sources.
Indeed, the solar system's orbital velocity around the galactic center 
causes an aberrational affect of the order of 2.5 ~arcmin. All measured 
star and quasar positions are shifted towards the point on 
the sky having galactic coordinates  $l=90^\circ, ~b=0^\circ$. 
For an arbitrary point on the sky the size of the effect 
is $2.5 ~\sin\eta$ ~arcmin, where $\eta$  is the angular distance to 
the point $l=90^\circ, ~b=0^\circ$. The acceleration of the solar system 
towards the galactic center causes this aberrational effect to change slowly.
This leads to a slow change of the apparent position of distant celestial
objects, i.e. to an apparent proper motion.

Let us assume a solar velocity of $220 $~km/sec and a distance of 8.5 ~kpc 
to the galactic center. The orbital period of the Sun is then 250 million years,
and the galactocentric acceleration takes a value of about 
$1.75 \times 10^{-13} $~km/sec$^2$. Expressed in a more useful units it is 
5.5 ~mm/s/yr. A change in velocity by 5.5 mm/sec causes a change in aberration
of the order of 4 $\mu$as. The apparent proper motion of a celestial object
caused by this effect always points towards the direction of the galactic center.
Its size is   $~4 \sin\eta ~\mu$as/yr, where $\eta$ is now the angular 
distance between the object and the galactic center.
 
The above hold in principle for quasars, for which it can be assumed that the 
intrinsic proper motions (i.e. those caused by real transverse motions) are 
negligible. A proper motion of   $4 ~\mu$as/yr corresponds to a transverse 
velocity of $2\times 10^4$~ km/sec at $z=0.3$ for $H_0$=100 ~km/sec/Mpc, and 
to $4\times 10^4 $~km/sec for $H_0=50 $~km/sec/Mpc.
Thus, all quasars will exhibit a distance-independent
steering motion towards the galactic center. 
Within the Galaxy, on the other hand, the effect is drowned in the local
kinematics: at 10 ~pc it corresponds to only 200 m/sec.

However, for a differential astrometry with SIM this effect will have to
be scaled down to account for the size of the field of 
regard \cite{aberr_memo}, namely
$2\sin\frac{\tt FoR}{2}=2\sin \frac{\pi}{24}=0.261$. This fact is reducing the
total effect of the galactocentric acceleration to only  $~\sim 1 \sin\eta
~\mu$as/yr ~and, thus, it makes   the detection of the solar system's
galactocentric acceleration  with SIM  to be a quite problematic issue.

\section{Deflection of Light by the Higher Multipoles
of the Gravity Field}
\label{sec:high_multipoles}

In order to carry out a  complete analysis of the  phenomenon of the 
relativistic light deflection one should account for other possible terms
in the expansion (\ref{eq:deflec0}) that may potentially contribute to this 
effect.   These terms are  due to non-sphericity and non-staticity of the
body's gravity field. 

\subsection{Gravitational Quadrupole Deflection of Light}

Effect of the gravitational deflection of light caused by the  quadrupole
term may be given  as \cite{Turyshev98}: 
{}
\begin{equation}
\theta_{J_2}= \frac{1}{2}(\gamma+1) \frac{4G M}{c^2{\cal R}}J_2 
\Big(1-s_z^2-2d_z^{\hskip 2pt 2} \Big)
\left({{\cal R}\over d}\right)^3,
\label{eq:quad}
\end{equation}
\noindent where  $J_2$ is the second zonal harmonic of the body under
question, 
$\vec{s}=(s_x,s_y,s_z)$ is the unit vector in the direction of the 
light ray propagation and vector $\vec{d}=d(d_x,d_y,d_z)$ is the 
impact parameter. A similar expression may be obtained for the
differential observations. This formula, for estimation purposes only, 
may be given as follows:
{}
\begin{equation}
\delta\theta_{J_2}\approx \frac{4\mu_B  J_{2_B}{\cal R}^2_B}
{r^3_B}
~\Big[\,\frac{1}{\sin^3\chi_{1B}}-\,\frac{1}{\sin^3\chi_{2B}} \Big].
\label{eq:quad_diff}
\end{equation}
\noindent The corresponding effects for the deflection of light by the
quadrupole mass distribution in the solar system planets  
are given in the Table \ref{tab11}. Note that the effect of the
quadrupole deflection of light depends on a number of different
instantaneous geometric parameters defining the mutual orientation of
the vector of the light propagation, position of the planet in orbit, 
the orientation of the axes defining it's figure, etc. This to model this
effect will require a significant effort. 

This effect depends  on the third power of the inverse impact parameter.
This fact together with a small planetary angular sizes (compare to the
size  of the SIM's field  of regard) makes it insensitive to the size of
the FoR for differential astrometry.  Note that the measurements of the
quadrupole deflection  have never been done before. SIM will allow to 
measure this effect directly for the first time. At the expected level
of accuracy the  knowledge of some  fundamental phenomena, such as the
jovian atmosphere, the magnetic field  fluctuations, etc.,  may
contribute to the errors in the experiment  \cite{TreuhaftLowe91}.  


\begin{table}[t]
\begin{center}
\caption{Higher gravitational coefficients for solar system bodies.   The data
taken from  {\tt http://nssdc.gsfc.nasa.gov/planetary/factsheet/}}  
\label{tab14}\vskip 10pt
\begin{tabular}{|r|c|c|c|} \hline
Solar system's & $J^B_2,$ & $J^B_4,$ & 
$J^B_6,$\\
object & $ \times ~10^{-6}$ & $\times ~10^{-6}$ & 
$\times ~10^{-6}$\\\hline\hline 

Sun & $0.17\pm0.017$& | & |  \\

Sun at 45$^\circ$ & | & |  & |  \\ 

Moon & 202.2  & $-0.1$   & |   \\

Mercury & 60.  &  | &  |  \\

Venus & 4.5  & $-2.1$ &  |  \\

Earth &  1,082.6  & $-1.6$ &  0.5 \\
 
Mars & 1,960.45 & | & | \\

Jupiter & 14,738$\pm$1 & $-$587$\pm$5 & 34$\pm$50\\ 

Saturn & 16,298$\pm 50$ & $-$915$\pm$80 & $103.0$\\ 

Uranus  & 3,343.43  & |   & |   \\

Neptune  & 3,411.  & | & |  \\
 
Pluto    &  |  &  | & |  \\\hline

\end{tabular}
\end{center}
\end{table} 
\begin{table}[h]
\begin{center}
\caption{Relativistic quadrupoles deflection of light by the 
bodies in the  solar system.} 
\label{tab11}  \vskip 10pt 
\begin{tabular}{|r||c|c|c||c|c|} \hline
Solar system' & $J^B_2,$ & $\theta^B_{J_2}, $ &
$d^{\tt crit}_{{J_2}}$   & $\delta\theta^B_{J_2}[15^\circ],$ & 
$\delta\theta^B_{J_2}[1^\circ],$\\
object & $ \times ~10^{-6}$ & $\mu$as &
 & $\mu$as & 
$\mu$as\\\hline\hline 

Sun & $0.17\pm0.017$& 0.3 & | & 0.3  & 0.3  \\


Moon & 202.2  & 2$\times 10^{-2}$ &   | 
              & 2$\times 10^{-2}$ & $2\times 10^{-2}$    \\

Mercury & 60.  & 5$\times 10^{-3}$  &  |&  | &  |   \\

Venus & 4.5  & 2$\times10^{-3}$ &  | & | &  |  \\

Earth &  1,082.6  & 0.6  & | & 0.6  &  0.6 \\
 
Mars & 1,960.45 & 0.2 & | & 0.2 & 0.2 \\\hline

Jupiter &  14,738$\pm$1  & 242.0 &  98$''$.12 \--- 144$''$.81  & 242.0  
& 242.0  \\
  &     &  &  6.23 ${\cal R}_J$  &    &  \\\hline


Saturn &  16,298$\pm$ 50  & 94.6 &  35$''$.62 \--- 43$''$.93 & 94.6 
& 94.6 \\ 
  &      & & 4.56 ${\cal R}_S$   &  &   \\\hline 

Uranus  & 3,343.43  & 7.3 &  3$''$.25 \--- 3$''$.61 & 7.3   & 7.3   \\
        &    &  & 1.94 ${\cal R}_U$  &     &    \\\hline

Neptune  & 3,411.   & 8.5 & 2$''$.23 \--- 2$''$.42 & 8.5 & 8.5  \\
  &    &  &  2.04~${\cal R}_N$ &  &    \\\hline
 
Pluto    &  |  &  |  & | &  | & |   \\\hline

\end{tabular}
\end{center}
\end{table}

As a result,  one will have to account for the quadrupole component of
the gravity fields  of the outer planets. In
addition, the influence of the  higher harmonic may be also of interest.
Let us estimate  the influence of some gravitational multipole
moments of Jupiter and Saturn, which are presented in the Table
\ref{tab11}.  It is convenient to discuss the  deflection by the
$J_2$ and $J_4$  coefficients  of the jovian gravity in terms of the
Jupiter-source separation angle $\chi_{1J}$.  
An  expression, similar to that of Eq.(\ref{eq:deflec_jup}) for the
monopole deflection, may be  obtained for the jovian quadrupole
deflection  in terms of the Jupiter-source separation angle $\chi_{1S}$.
The quadrupole deflection angle in this case may be given  as:
{} 
\begin{equation}
\theta^{\tt max}_{J_2}={3.46058\times 10^{-10}}~
\frac{1}{\sin^3\chi_{1J}} ~~\mu{\sf as}.
\label{(2.16)}
\end{equation}

\noindent 
The Jupiter's angular dimensions   
from the Earth are  calculated to be ${\cal R}_J=23.24 ~$arcsec, 
which correspond to a deflection  angle of $242 ~\mu$as. 
The deflection on the multipoles  for some   $\chi_{1J}$ is given  in
the Table \ref{tab12}. 


\begin{table}[h]
\begin{center}
\caption{Deflection of light by the Jovian higher gravitational
coefficients.} 
\label{tab12}\vskip 10pt
\begin{tabular}{|r|c|c|c|c|c|c|c|}
\hline

\tt Jovian   & \multicolumn{7}{c|}{$\chi_{1J}$, arcsec}   
\\ \cline{2-8} 
\tt deflection &$23''.24$ &26$''$&30$''$&35$''$&40$''$&50$''$
&120$''$\\\hline\hline
$\theta^J_{J_2}, ~\mu$as &
242 &
173 & 
112 &
71 &
47 & 
24 &
1.8 \\ \hline

$\delta\theta^J_{J_2}[15^\circ], ~\mu$as &
242 &
173 & 
112 &
71 &
47 & 
24 &
1.8  \\ \hline\hline

$\theta^J_{J_4}, ~\mu$as &
9.6 &
5.5 & 
2.7 &
1.3 &
0.6 & 
0.2 &
0.0  \\ \hline
\end{tabular} 
\end{center}
\vskip 10pt
\begin{center}
\caption{Deflection of light by the Saturnian higher gravitational
coefficients.} 
\label{tab12sat}\vskip 10pt
\begin{tabular}{|r|c|c|c|c|c|c|c|}
\hline

\tt Saturnian   & \multicolumn{7}{c|}{$\chi_{1S}$, arcsec} \\
\cline{2-8} 
\tt deflection &$9''.64$ &12$''$&15$''$&20$''$&25$''$&30$''$&35$''$
\\\hline\hline
$\theta^S_{J_2}, ~\mu$as &
94.7 &
49.1 & 
25.1 &
10.6 &
5.4 & 
3.1 &
2 \\ \hline
$\delta \theta^S_{J_2}[15^\circ], ~\mu$as &
94.7 &
49.1 & 
25.1 &
10.6 &
5.4 & 
3.1 &
2 \\ \hline\hline

$\theta^S_{J_4}, ~\mu$as &
5.3 &
1.8 & 
0.6 &
0.1 &
| & 
| &
| \\ \hline

\end{tabular} 
\end{center}
\end{table}

A similar studies could be performed for the Saturn.
In terms of the Saturn-source separation angle $\chi_{1S}$
the saturnian quadrupole  deflection mat be estimated with the help of
the following expression: 
{}
\begin{equation}
 \theta^{\tt max}_{J_2}=9.66338\times 10^{-12}~
\frac{1}{\sin^3\chi_{1S}} ~~\mu{\sf as}.
\label{eq:guad_sat}
\end{equation}
The Saturn's angular dimensions  from the Earth' orbit  are  calculated
to be
${\cal R}_S=9.64 ~$arcsec, which correspond to a deflection  angle of
$94.7~\mu$as.  The corresponding estimates for the deflection angles 
are presented in the Table \ref{tab12sat}.

As a result, for  astronomical observations with  accuracy 
of about   $1 ~\mu{\rm as}$, 
one will have to account for the quadrupole gravitational fields 
of the Sun, Jupiter, 
Saturn, Neptune, and Uranus. In addition, the influence of the 
higher harmonics may be of interest.
For example some of the moments for Jupiter and Saturn are given 
in the Table \ref{tab14}.
Concluding this paragraph, we would like to note that the higher
multipoles   may also influence the astrometric observations taken close
to these planets. Thus, for both Jupiter and Saturn the  rays, grazing
their surface, will  be deflected by the fourth zonal harmonic $J_4$ as
follows: 
$\delta\theta^J_{J_4}\approx  9.6 ~\mu$as, 
$\delta\theta^S_{J_4}\approx  5.3 ~\mu$as. 
In addition, the contribution of the $J_6$ for Jupiter and Saturn will
deflect the grazing rays  on the angles   
$\delta\theta^J_{J_6}\approx
0.8 ~\mu$as, $\delta\theta^S_{J_6}\approx 0.6 ~\mu$as. The contribution 
of $J_4$ is decreasing with the distance from the body as $d^{-5}$ and
contribution of $J_6$  as $d^{-7}$. As a result the deflection angle will
be less then  $1 ~\mu$as when $\hskip 2pt d>1.6 ~{\cal R}$, where 
${\cal R}$ is the radius of the planet.

\subsubsection{Critical Distances for Quadrupole Deflection  of Light}

The critical distance $d^{\tt crit}_{J_2}$ for the  astrometric
observations in the regime of quadrupole deflection of light 
with accuracy of $\Delta \theta_0= \Delta k ~\mu$as  was defined  as:   
{}
\begin{equation}
d^{\tt crit}_{J_2} = {\cal R}_B \Big[\frac{4\mu_B}{{\cal R}_B}  
\frac{J^B_2}{\Delta\theta_0}\Big]^\frac{1}{3}.
\label{eq:quad_crit_abs}
\end{equation}
\noindent 
The  critical distances for the relativistic quadrupole  deflection of
light by the solar system's bodies   for the case of $\Delta
k=1$ presented   in the Table
\ref{tab11}.

\subsection{Gravito-Magnetic Deflection of Light}

Besides the gravitational deflection of light by the monopole and the
quadrupole components of the  static gravity filed of the bodies, the
light ray  trajectories will  also  be affected by the non-static
contributions from this field.  It is easy to demonstrate  that a
rotational motion of a  gravitating  body contributes to the total
curvature of the space-time generated by this same body. This
contribution produces an  additional deflection of light rays on the
angle  {}
\begin{equation}
\delta\theta_{\vec{\cal S}}=\frac{1}{2}(\gamma +1)
\frac{4G}{c^3d^3}\vec{\cal S}  (\vec{s}\cdot\vec{d}),
\label{eq:defl_spin}
\end{equation} 
\noindent where ${\vec{\cal S}}$  is the body's angular momentum. 

The most significant contributions of  rotation of the 
solar system bodies  to the relativistic  light deflection
are the following ones:  the solar deflection amounts to
$\delta\theta^\odot_{\vec{\cal S}}=\pm(0.7~-1.3)\mu$as 
[the first term listed is for a uniformly rotating Sun; 
the second is for the Dicke's model]; jovian is about
$\delta\theta^J_{\vec{\cal S}}=\pm0.2~\mu$as;  and saturnian
$\delta\theta^{Sa}_{\vec{\cal S}}=\pm0.04~\mu$as.  
Thus, depending on the model for the solar interior,    
solar rotation  may 
produce a noticeable contribution for the grazing rays.  
The estimates of magnitude of deflection of light ray's trajectory, 
caused by the rotation of gravitating  bodies demonstrate that for 
precision of observations of $1 ~\mu$as it is sufficient to account 
for influence of the Sun and Jupiter only.

The relativistic gravito-magnetic deflection of light 
has never been tested before. Due to the fact that  the magnitudes of 
corresponding  effects in the solar system 
are too small and, moreover,  the  SIM operational mode 
limits the viewing angle for a sources as $\chi_{1\odot}\ge45^\circ$,
SIM will not be sensitive to this effect.

\section*{Discussion}

General relativistic deflection of light  produces  a significant
contribution to the future astrometric observations with accuracy of
about a few $\mu$as.  In this Memo we addressed the problem of light
propagation on the gravitational field of the solar system. It was shown
that for high accuracy observations   it is necessary to correct for the
post-Newtonian deflection of  light by  the monopole  components of  
gravitational fields of a large number of celestial bodies in the solar
system, namely the Sun and the nine planets, together with the planetary
satellites and the  largest asteroids (important only if observations
are conducted in their close proximity). The most  important fact
is that the gravitational presence of the Sun,  the Jupiter and the 
Earth  should  be always taken into  account,   independently on the
positions of these bodies relative to the interferometer. It is worth
noting that the post-post Newtonian effects due to the solar gravity
will not  be  accessible with  SIM. This effect as well as the effect of
gravitational deflection of light caused by the  mass quadrupole term of
the Sun are negligible at the  level of expected  accuracy. However,
deflection of light by some planetary  quadrupoles may have a big
impact on the astrometric accuracy. Thus, the higher
gravitational multipoles should be taken into account when 
observations are conducted in the close proximity of two bodies
of the solar system, notably the Jupiter and the Saturn.

We addressed the  problem of adequacy of the current level of
accuracy of the solar system ephemerides. It turns out that, even though
the accuracy in determining the outer planets positions is below the
general relativistic requirements, one may expect that SIM will actually
improve the planetary ephemerides  simply as a by-product of its 
future astrometric campaign. 

As an important  result of it's astrometric campaign, SIM could provide an
accurate  measurement  of the PPN parameter $\gamma$. Thus, for observations on
the rim of the solar avoidance angle one  could  determine $\gamma$ to an
accuracy  of about two parts in $10^3$ in a single measurement.     For a large
number of observing pairs of stars   such an experiment
could potentially determine $\gamma$  with an accuracy of about
$\sim 10^{-5}$ which  is an order of magnitude better than
presently known.  One could perform experiments with a comparable
accuracy in the Jupiter's gravity field.  To correctly address this
problem an extensive covariance studies are needed.

\acknowledgments
The reported research   has been done at the Jet Propulsion
Laboratory,  California Institute of Technology, which is under  contract to the 
National Aeronautic and Space Administration.

\clearpage

\begin{table}
\begin{center}
\caption{Some astronomical parameters for the bodies in our  solar system.
\label{tab:solar_system_param}}\vskip 10pt
\begin{tabular}{|r|r|r|r|r|}\hline
Object & Mean distance,  & Radius  & Inverse mass, & Sidereal\\ 
&  AU ~(1900.0) & ${\cal R}_B$,  km & ${\cal M}_\odot/{\cal M}_p$ & 
period, yr\\
\hline
\hline

       Sun & 8.5 ~kpc    & 695,980  & 1.00    & $2.5\times 10^8$\\  
      Moon & 384,400~km  & 1,738    & 27,069,696.00 & 2 \\  
   Mercury & 0.3870984   & 2,439    & 6,023,600.00  & 0.241  \\ 
     Venus & 0.7233299   & 6,050    & 408,523.71    & 0.615  \\  
     Earth & 1.0000038   & 6,378.16 & 332,946.05    & 1.000  \\  
      Mars & 1.5237      & 3,394    & 3,098,708.00  & 1.881  \\  
   Jupiter & 5.2037      & 70,850   & 1,047.35      & 11.865 \\  
    Saturn & 9.5803      & 60,000   & 3,497.99      & 29.650 \\  
    Uranus & 19.1410     & 24,500   & 22,902.98     & 83.744 \\  
   Neptune & 30.1982     & 25,100   & 19,414.24     &165.510 \\  
    Pluto  & 39.4387     & 3,200    
                         & $1.35 \times 10^8$ & 247.687 \\ \hline

\end{tabular}
\end{center}
\vskip 40pt  
\begin{center}
\caption{Some physical constants and conversion factors used 
in the paper.
\label{tab:constants}}\vskip 10pt
\begin{tabular}{r l} 
 
Relativity constant: & $G/c^2=0.7425\times 10^{-28}~~$cm/g, \\
Speed of light: &    $c = 2.997292\times 10^{10}~~$cm/sec,  \\
Solar mass: & ${\cal M}_\odot= 1.9889\times 10^{33}~~$g,  \\  
Solar gravitational constant: & 
$\mu_\odot  = c^{-2}G {\cal M}_\odot=1.47676\times 10^{5} ~~$cm,  \\ 
Solar quadrupole coeff.: & $J_{2\odot} = (1.7\pm0.17)\times 10^{-7}$, 
\\   Solar spin moment: &  ${\cal S}_\odot =  1.63\times
10^{48}~~$g~cm$^2$/sec,  \\ 
Earth|Moon distance: & $r_{\oplus-m} 
 = 3.844\times 10^{10}~~$cm, \\
Earth's spin moment: & 
            ${\cal S}_\oplus=5.9\times 10^{40}~~$g~cm$^2$/sec,\\
Astronomical Unit:& AU = $1.495~978~92(1)\times 10^{13}~~$cm,  \\ 
1 parsec: & pc $=3.0856\times 10^{18}~~$cm, \\
1 light-year: & ly $=0.94605\times 10^{18}~~$cm, \\
1 year: & yr $=3.155~692~6\times 10^{7}~~$sec, \\
1 day: & day $=86~400~~$sec,\\
1 sidereal day: & s$\_$day $=86~164.091~~$sec, \\
1 microarcsecond: & $1~\mu$as $=4.84814\times 10^{-12} ~~$rad, \\ 
1 radian: & 1~rad $=0.20627\times 10^{12} ~~\mu$as.\\ 
\end{tabular}
\end{center}
\end{table}


\newpage
\begin{thebibliography}{31}

\bibitem[Perryman et al. (1992)]{Perryman92}  Perryman, M. A. C.,
 et al.  1992,  \aap,   258, 1 

\bibitem[Gould (1993)]{Gould93} Gould, A. 1993, \apj,  414, L37 
 
\bibitem[Turyshev \& Unwin (1998)]
{aberr_memo} Turyshev, S. G. and
Unwin, S. C. 1998,  
     {\it Relativistic Stellar Aberration Requirements for  
     the Space Interferometry Mission},  
 JPL Technical Memorandum \#98-1017, Pasadena, CA.

\bibitem[Sovers \& Jacobs (1996)]{Modest96} Sovers, O. J., Jacobs, C.
S. 1996, in  {\it Observation Model and Parameter Partials for the JPL VLBI
Parameter  Estimation Software "MODEST" - 1996}, JPL
Technical  Report 83-39, Rev. 6, Pasadena, CA.


\bibitem[Will (1993)]
{Will93} Will, C. M. 1993, 
{\it Theory and Experiment in Gravitational Physics}, 
     (Rev. Ed.), Cambridge  Univ. Press, Cambridge, England.
 

\bibitem[Dar (1992)]{Dar92}
     Dar, A. 1992, {\it Nucl. Phys.}, {\bf B} (Suppl.), 28A, 321

\bibitem[Treuhaft \&  Lowe (1991)]
{TreuhaftLowe91}
    Treuhaft, R. N.,  \& S. T. Lowe: 1991, 
\aj, 102, 1879

\bibitem[Eubanks et al (1997)]
{Eubanks97}
     Eubanks, T. M.  et al.: 1997 
     ``Advances in Solar System Tests of Gravity.'' 
     In: Proc. of The Joint
     APS/AAPT 1997 Meeting, 18-21 April 1997, Washington D.C. 
     {\em Bull. Am. Phys. Soc.}, 
     Abstract \#K 11.05 (1997), unpublished.


\bibitem[Turyshev (1998)]
{Turyshev98} Turyshev, S. G. 1998,
\baas, 29, 1223

\bibitem[Jacobs et al (1998)]
{Sovers98}
 Sovers, O. J.,   Fanselow, J. L., and  Jacobs, C. S. 1998,
 70, 1393


\bibitem [Standish \& Hellings (1989)]
{StandishHellings89}
    Standish, E. M. Jr.,  Hellings, R. W.  
    1989, {\it Icarus},   80, 326

\bibitem[Yoder (1995)]
{Yoder95}
Yoder, C F. 1995,
{\it Astrometric and Geodetic Properties of Earth and the Solar System}.
Global Earth Physics. A Handbook of Physical Constants, 
AGU Reference Shelf 1.

\bibitem[Standish (1995)]
{Standish95}
 Standish, E. M. Jr. 1995,  {\it Astronomical and Astrophysical 
     Objectives of Sub-Milliarcsecond Optical Astrometry. IAU-SYMP}, 
    {\bf166}, 
     eds. E. H\"{o}g and P. K. Seidelmann. p.109 


\bibitem[Standish et al (1995)]
{Standish_etal95}
     Standish, E. M. Jr.,  Newhall, X X, Williams, J. G., 
     and Folkner, W. M. 1995,
     {\it JPL Planetary and Lunar Ephemeris, DE403/LE403}, 
     Jet Propulsion Laboratory  IOM \# 314.10-127 

\bibitem[Folkner  et al (1994)]
{Folkner94}
    Folkner, W. M. {\it et al.}  1994, 
    \aap, 287, 279   

\end{thebibliography}
\end{document}